\documentclass[11pt,onecolumn,journal,cspaper,draftcls]{IEEEtran}
\def\baselinestretch{1.4}

\usepackage{graphicx,color,psfrag}
\usepackage{amsmath,amsfonts,amssymb}
\usepackage{multirow}
\usepackage{booktabs}
\usepackage{pstricks}
\usepackage{cite}

\newcommand{\supscr}[1]{\ensuremath{^{\textrm{#1}}}}
\newcommand{\subscr}[1]{\ensuremath{_{\textrm{#1}}}}

\def\Foursquare{\textsf{Foursquare}}
\def\Gowalla{\textsf{Gowalla}}
\def\GLatitude{\textsf{Google Latitude}}
\def \Flickr {\textsf{Flickr}}
\def\Facebook{\textsf{Facebook}}

\begin{document}

% Title portion
\title{Personalised Travel Recommendation based on Location Co-occurrence}

\author{Maarten~Clements,
				Pavel~Serdyukov,
				Arjen~P.~de~Vries,
				Marcel~J.T.~Reinders% <-this % stops a space
\thanks{M.~Clements is with TomTom., Netherlands. E-mail: maarten.clements@gmail.com.
P.~Serdyukov is with Yandex. E-mail: pavser@yandex-team.ru.
A.P.~de~Vries is with National Research Institute for Mathematics and Computer Science (CWI). E-mail: arjen@acm.org.
M.J.T.~Reinders is with Delft University of Technology. E-mail: m.j.t.reinders@tudelft.nl.\protect\\\protect\\
\noindent\textbf{This work has been submitted to the IEEE for possible publication. Copyright may be transferred without notice, after
which this version may no longer be accessible.}}}

\markboth{Submitted to IEEE Transactions on Knowledge and Data Engineering, June~2011}%
{Clements \MakeLowercase{\textit{et al.}}: Personalised Travel Recommendation based on Location Co-occurrence}

\IEEEcompsoctitleabstractindextext{%
\begin{abstract}
We propose a new task of recommending touristic locations based on a user's visiting history in a geographically remote region. This can be used to plan a touristic visit to a new city or country, or by travel agencies to provide personalised travel deals. 

A set of geotags is used to compute a location similarity model between two different regions. The similarity between two landmarks is derived from the number of users that have visited both places, using a Gaussian density estimation of the co-occurrence space of location visits to cluster related geotags. The standard deviation of the kernel can be used as a scale parameter that determines the size of the recommended landmarks. 

A personalised recommendation based on the location similarity model is evaluated on city and country scale and is able to outperform a location ranking based on popularity. Especially when a tourist filter based on visit duration is enforced, the prediction can be accurately adapted to the preference of the user. An extensive evaluation based on manual annotations shows that more strict ranking methods like cosine similarity and a proposed RankDiff algorithm provide more serendipitous recommendations and are able to link similar locations on opposite sides of the world.
\end{abstract}

%\category{H.3.5}{Information Storage and Retrieval}{Online Information Services}
%\category{H.3.3}{Information Storage and Retrieval}{Information Search and Retrieval --- Information filtering}
%\category{H.2.8}{Database Management}{Database Applications --- Data mining}
%\category{H.2.8}{Database Management}{Database Applications --- Spatial databases and GIS}
%\terms{Algorithms, Experimentation, Human Factors}
%\keywords{Personalization, Location Based Services, Recommendation, Social Media, Flickr, Geotag} % NOT required for Proceedings
}

\maketitle

\IEEEdisplaynotcompsoctitleabstractindextext

\IEEEpeerreviewmaketitle

\section{Travel Recommendation}
%
%$2 \cdot 10^{-3}\,\textup{kg}$
%or siunitx
%(http://www.latex-community.org/forum/viewtopic.php?f=46&t=3684)

Location based services are quickly gaining popularity due to affordable mobile devices and ubiquitous Internet access. Websites like \Foursquare\footnote{http://foursquare.com/}, \Gowalla\footnote{http://gowalla.com/}, \GLatitude\footnote{http://www.google.com/latitude} and \Facebook\footnote{http://www.facebook.com/} show that people want to share their location information and get accurate location recommendations at any time and and place. In return for sharing their location data, users can now be matched to products, venues, events or local social relations and groups. 

Accurate predictions of the user's preferred locations can simultaneously aid the user itself, advertisers of products specific to the recommended place and service providers (e.g.\;transportation to the recommended location). To provide these recommendations, the system needs to have an accurate way to find similarities between locations or people. We propose to exploit the past visiting behaviour of people to build a location similarity model that can be used for personalised location predictions. 

In this work we will exploit a set of \textit{geotags} collected from \Flickr\footnote{http://flickr.com/} to make a recommender that can predict relevant locations for individual users. In \Flickr, geotags are tuples of latitude and longitude that represent the exact location where a user made a photo. Registration of geotags can be done manually by placing the photo on a map, or automatically by the device if it is equipped with a GPS module. Here we show that the collective knowledge represented in these geotags can be used to estimate similarities between locations and that personalised location recommendations can be derived from this similarity model. 

Given a user's preference in one predefined area, we predict his activity in a another disjoint area. The proposed method will be evaluated on both city and country scale and will show that places on opposite sides of the world can be related based on user location histories. 

\section{Related Work}
Since GPS equipped mobile phones have become mainstream, the amount of available geotags has grown to a number that allows for intensive data analysis. In this work, geotags are used to predict interesting locations for individual users, but the exploitation of geotags has shown to be effective for various other tasks. A method for global event detection has been proposed by Rattenbury et al., who searched for the occurrence of textual tags in spatial and temporal bursts~\cite{Rattenbury2007}. Ahern et al.\ made a mapping of popular tags to geographical locations, resulting in a scale dependent map overlay with semantic information on the underlying data~\cite{Ahern2007}. This work was extended by Kennedy et al.\ who selected relevant pictures for the predicted clusters~\cite{Kennedy2007}. Crandall et al.\ suggested not to use a fixed number of clusters and proposed a mean shift algorithm to find the most prominent landmarks and representative photos~\cite{Crandall2009}. 

Another application of \Flickr's geotags was proposed by Lee et al.\ who used the geographical clusters related to a tag to improve the prediction of similar tags~\cite{Lee2008}. Furthermore, several methods have been proposed to predict the geotags of a photo, based on its textual tags~\cite{Serdyukov2009}, visual information~\cite{Crandall2009} and individual user travel patterns~\cite{Kalogerakis2009}.

As geotags relate to a location where the user made a photo, they inherently contain a touristic preference indication. Full GPS tracks are useful to study daily mobility patterns but extra effort is needed to extract touristically interesting spots. Based on users' GPS tracks, location recommender systems have been proposed that attempt to predict popular places and activities near the current location of the user. Some work has focused on the recommendation of specific types of locations. An item-based collaborative filtering method was used to recommend shops, similar to a user's previously visited shops~\cite{Takeuchi2006} and a user-based collaborative filtering was proposed to generate restaurant recommendations through users with similar taste~\cite{Horozov2006}. Zheng et al.\ extensively studied GPS tracks in Beijing, defined a method to extract interesting locations from this data (\textit{Stay regions}) and proposed a matrix factorization method to suggest locations and activities based on the current state of the user~\cite{Zheng2010}. They also showed that the HITS model can effectively be used to create a ranking of popular locations and experienced people~\cite{Zheng2009a}.   

Compared to most of the previously proposed methods, our system gives recommendations in a geographically remote location, so people can use it when they are planning a trip to another country or city. We have previously showed that geotags can be used to construct a measure of similarity between locations~\cite{Clements2010}. Here, we present a thorough extension of the previous work, using a similarity model based on a scale-space of location co-occurrence data. We evaluate the potential of this similarity model for personalised recommendations. The proposed model contains a scale parameter that allows the prediction of differently sized regions. So, when a user decides to visit a certain country the recommender can be used to find the most interesting cities and when a user gets to that city the same method can be used to find the most interesting landmarks, restaurants or other venues. 

Many recommendation algorithms have been proposed based on similarities between objects in a discrete item-space~\cite{Sarwar2001,Wang2006c}, which has proven to be effective in E-commerce applications~\cite{Linden2003}. Compared to these systems, a location recommender does not have a limited number of objects to recommend. Any point consisting of two continuous values of latitude and longitude can be recommended. On a more fundamental level, we introduce a model that includes the pairwise distances between points in order to reason in this continuous space. We will demonstrate the effectiveness of this model on geographical data, but it could easily be extended to include other continuous dimensions like temporal information. 

\begin{figure}[!t]
\begin{minipage}[b]{0.51\columnwidth}%
\centering
\includegraphics[width=\columnwidth]{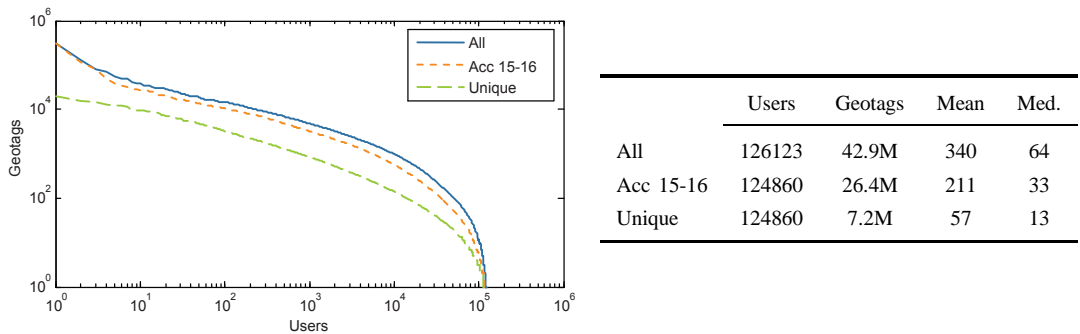}
\end{minipage}
\hspace{0pt}
\begin{minipage}[b]{0.47\columnwidth}%
\scriptsize
\begin{tabular}{lcccc}
\toprule
& Users & Geotags & Mean & Med.\\
\cmidrule{2-5}
All & 126123 & 42.9M & 340 & 64\\
Acc 15-16 & 124860 & 26.4M & 211 & 33\\
Unique & 124860 & 7.2M & 57 & 13\\
\bottomrule
\end{tabular}
\vspace{35pt}
\end{minipage}
\caption{The distribution of the number of geotagged photos per user in descending order. The accuracy filter reduces the data set from 43M to 26M geotags. By selecting only unique geotags we maintain 7M points. The table also indicates the mean and median number of geotags per user.}
\label{fig:DataStats}
\end{figure}

\section{Data}\label{sec:DataFlickr}
\subsection{Data Collection}

%Describe Flickr crawl? - First collect seed cities, Get a photo, get this user's photos. 
Using the public API of \Flickr\ we have collected a large set of geotagged photos in a period of several months at the end of 2009 and early 2010. Figure~\ref{fig:DataStats} gives the distribution of the number of geotags per user (\textit{All}). The distribution clearly shows that our crawl has a bias to people with many geotags, as the expected long tail of the distribution is missing. However, as we will only evaluate recommendations for users who have provided a sufficient amount of data, this bias in the crawl does not interfere with the objectives of this work. The total set corresponds to roughly 46\% of the 93 million publicly available geotags in \Flickr\ at the end of 2009\footnote{According to: http://www.flickr.com/map}.

Each geotag has an associated level of accuracy in the range of 1-16, 16 being the most accurate. This accuracy roughly relates to the zoom level of the map interface in \Flickr. Because we want to make accurate predictions at the scale of individual landmarks, we keep only geotags at accuracy 15 or 16 (street level). The remaining data is represented by \textit{Acc 15-16} in Figure~\ref{fig:DataStats}. The possibility to integrate the accuracy value in the recommender system will be discussed in Section~\ref{sec:Discussion}.

\Flickr\ allows users to upload and annotate photos in \textit{batches}. When someone uses this function it can either mean that he made many photos at that location, or that he did not take the effort to give the exact coordinates for each individual photo. Because of the uncertainty about the user's intent when uploading a batch to a single location, we choose to ignore the possible relation between user preference and batch size and store only one geotag per batch. After these filtering steps, we retain 7.2 million geotags contributed by 125 thousand users (\textit{Unique} in Figure~\ref{fig:DataStats}). 

%===============================================================
% BEGIN FIGURE =================================================
\begin{figure}[!t]
\centering
\includegraphics[width=\columnwidth]{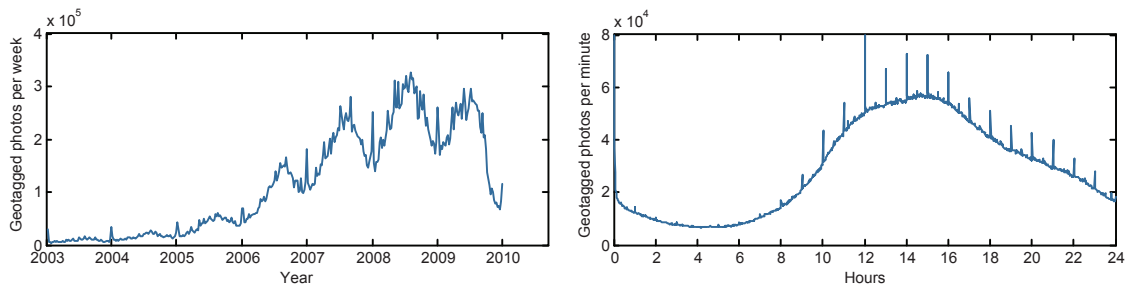}
\caption{When \Flickr\ users make photos. Left: Photo count per week from 2003 to 2010. Right: Photo count per minute of the day, aggregated over all days.}
\label{fig:DataTime}
\end{figure}
% END FIGURE ==================================================
%==============================================================

\subsection{Data statistics}
The collected data set gives an interesting insight in the common behaviour of \Flickr\ users. Besides the location of photos, \Flickr\ also stores the date and time a photo was taken (according to the internal camera clock). Figure~\ref{fig:DataTime} shows the number of photos taken in a certain week between 2003 and 2010. Apart from the clear increase in popularity over the last 5 years it is interesting to see that most of the photos are taken during the northern hemisphere summer. 

When we aggregate over all days and count the number of photos for each minute, we clearly see the bulk of photos is made late in the morning or early afternoon. In the evening the number of photos slowly decays until the minimum is reached around 4:30. The spikes at full hours and at January 1\supscr{st}  in the weekly histogram are caused by default values of empty fields in \Flickr's database. 

\begin{figure}[!b]
\centering
\includegraphics[width=0.8\columnwidth]{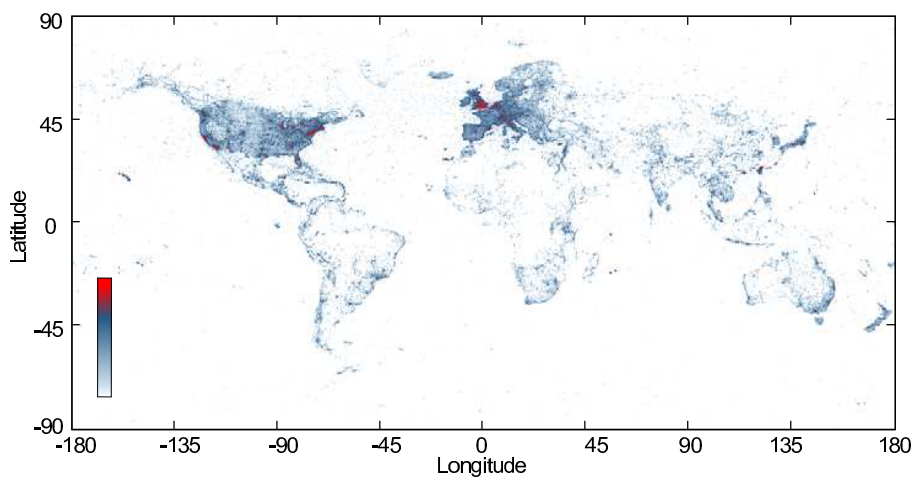}
\caption{Where \Flickr\ users make photos: World distribution.}
\label{fig:DataHist1}
\end{figure}

\begin{figure}[!t]
\centering
\includegraphics[width=0.8\columnwidth]{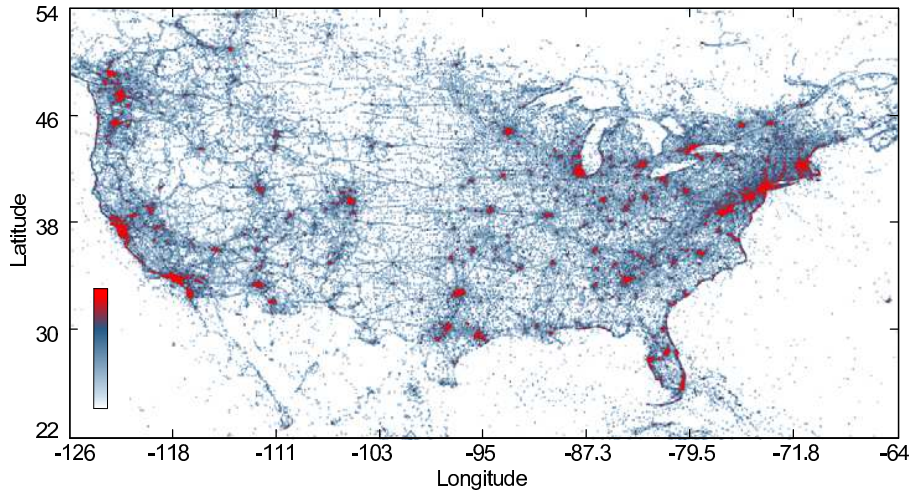}
\caption{Where \Flickr\ users make photos: USA distribution}
\label{fig:DataHist2}
\end{figure}

Figure~\ref{fig:DataHist1} gives the geographical distribution of the data. This 2000x4000 histogram of the geotags clearly shows the most popular travel areas in the \Flickr\ community. Europe and North America have the largest density of data points, but the rest of the world is also recognisable. Figure~\ref{fig:DataHist2} gives a closer view of North America, which shows that coastlines, cities and even highways are clearly represented in the data.

%===============================================================
% BEGIN FIGURE =================================================
% END FIGURE ==================================================
%==============================================================

%===============================================================
% BEGIN FIGURE =================================================
% END FIGURE ==================================================
%==============================================================

Based on this data, we select the 10 most popular countries and 10 most popular cities to evaluate the feasibility of personalised travel recommendation. We rank the cities and countries by the number of users that have been there (Table~\ref{tab:UserCount}), based on their geotags located within city bounding boxes\footnote{Collected in January 2010 from http://developer.yahoo.com/geo/geoplanet/} and country polygons\footnote{Collected in March 2010 from http://mappinghacks.com/}. Because the number of users in the USA is much larger than other countries, we split the USA in 3 regions: East USA (Longitude $>-98.583^\circ$), West USA (Longitude $<-98.583^\circ$), Alaska (Latitude $>50^\circ$). 

\begin{table}[t]
\caption{Number of users in top-10 cities and countries}
\label{tab:UserCount}
\centering
\footnotesize
\begin{tabular}{llll}
\toprule
Users&City&Users&Country\\
\cmidrule(r{.25em}){1-2}\cmidrule(l{.25em}){3-4}
19802 &	London, England, United Kingdom&45738&United States EAST\\
18291 &	New York, NY, United States&32904&United States WEST\\
13786 &	Paris, Ile-de-France, France&25934&United Kingdom\\
12470 &	San Francisco, California, United States&18247&France\\
7893 &	Rome, Lazio, Italy&16995&Italy\\
7627 &	Los Angeles, California, United States&15414&Spain\\
7208 &	Washington, District of Columbia, United States&13381&Germany\\
7158 &	Chicago, Illinois, United States&11024&Canada\\
7069 &	Barcelona, Catalonia, Spain&6503&Netherlands\\
6569 &	Berlin, BE, Germany&5067&Australia\\
\bottomrule
\end{tabular}
\end{table}

\section{Experimental setup}\label{sec:setup}

\begin{figure}[!t]
%\rput[tl](1.6,4.1) {\Large $U$}
%\rput[tl](3.3,6.5) {\Large $\mathcal{L}$}
%\rput[tl](3,1.6) {\Large $u_2$}
%\rput[tl](7.1,1.6) {\Large $\Phi$}
%\rput[tl](7.4,6.8) {\Large $l \in C$}
%\rput[tl](7.7,3.6) {\Large $f_{A}\ast g$}
%\rput[tl](7.7,3.6) {\large Conv.}
%\rput[tl](8.8,6.2) {\Large $u_1$}
%\rput[tl](10.8,4.1) {\Large $\Phi_{u_1}$}
%\rput[tl](10.8,2.2) {\Large $\Phi^{CC}$}
%\rput[tl](11.4,5.4) {\Large $f_{u_1}\ast g$}
%\rput[tl](11.4,5.4) {\large Conv.}
\begin{minipage}[b]{\columnwidth}%
	\begin{center}
	\includegraphics[width=0.6\columnwidth]{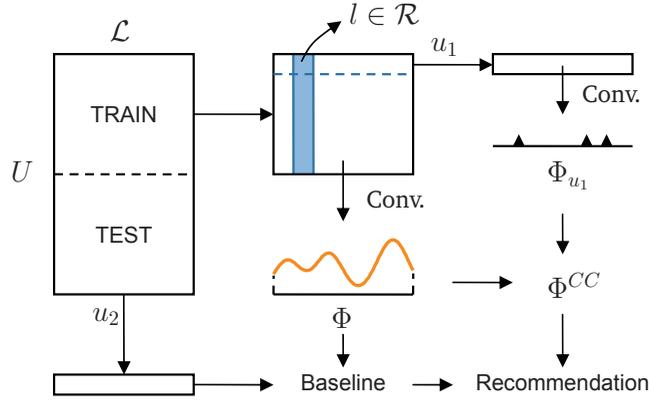}
	\end{center}%
\end{minipage}
\caption{Experimental setup. The training users generate the global travel distribution $\Phi$ and the location similarity model $\Phi^{CC}$. The performance of both models for location recommendation in a predefined region $\mathcal{R}$ is evaluated on the test users.}
\label{fig:DataFlow}
\end{figure}

Figure~\ref{fig:DataFlow} presents the experimental setup and the notation described in the following sections is summarised in Table~\ref{tab:Notation}. The data is comprised of a set of users $u \in U$ who have all visited at least one location $l \in \mathcal{L}$, where $l$ is a tuple $(x,y,z)$ of Cartesian coordinates and $\mathcal{L} \subset \mathbb{R}^3$ is the set of all geotags in our data set. The set of geotags $\mathcal{L}$ is a subset of the world $\mathcal{W}$ described by a sphere with radius 6,367,449\,m centered at zero. While \Flickr\ provides the geotags in latitude and longitude we will use Cartesian coordinates throughout this work, which is more efficient for the computation of Euclidean distances between points. The distance between two points is measured through the crust of the Earth instead of over the surface. This difference is negligible for small distances and rank equal in general. 

The data from half of the users (the \textit{training} set) will be combined in a model that captures the similarities between the most important locations in two regions. With the data from the other half of the users (the \textit{test} set) the application of the learned co-occurrence model for personalized travel recommendations will be evaluated.
%Half of the data will be used to compute the similarities between the most important locations in different regions (the \textit{training} set). The other half of the data (the \textit{test} set) will be used to evaluate whether the learned co-occurrence model can effectively be applied for personalized travel recommendations. 
We split the data in equally sized training and test sets by first ranking all users according to the number of geotags. In this order, we select users $1,4,5,8,9,\ldots$ as training users and $2,3,6,7,10,\ldots$ as test users, so the two sets will roughly follow the same distribution. 

\begin{table}[!b]
\caption{Notation used in this paper. For all $l$, $\mathcal{L}$, $f$, $\Phi$, $p$, $\mathcal{P}$ we use the superscript $\ldots^{s/t}$ to refer to the region of the data ($\mathcal{R}^s$ or $\mathcal{R}^t$) and the subscript $\ldots_{u_k}$ if the data is based on a single user. %For $p$ and $\mathcal{P}$, we also allow the subscript $\ldots_{U}$ to indicate the collection of peaks aggregated over all users. 
The locations in the co-occurrence space ($c$, $\mathcal{C}$) can also contain the subscript $\ldots_{u_k}$, but no superscript.}
\label{tab:Notation}
\centering
\footnotesize
\begin{tabular}{ll}
\toprule
$u_k \in U$    & The users in the \Flickr\ data set\\
$\mathcal{W}$  & The world; subspace of $\mathbb{R}^3$\\
$\mathcal{R}^s$, $\mathcal{R}^t$   & Starting region, target region; Subspaces of $\mathcal{W}$\\
$l \in \mathcal{L}$  & All geotags in the data set, subset of $\mathcal{W}$\\
%$l_{u_k} \in \mathcal{L}_{u_k}$  & The geotags of $u_k$\\
%$l^{s/t} \in \mathcal{L}^{s/t}$  & The geotags within $\mathcal{R}^s$ or $\mathcal{R}^t$\\
$f$  & Function describing a set of geotags\\
$\Phi$ & Function describing the Gaussian convolution of $f$\\
$p \in \mathcal{P}$  & The peaks of $\Phi$\\
$c \in \mathcal{C}$  & Points in the co-occurrence space; Subset of $\mathbb{R}^6$\\
\bottomrule
\end{tabular}
\end{table}

The objective of this work is to predict the visited locations of a test user $u_k \in U$ in a target region $\mathcal{R}^t \subset \mathcal{W}$, based on the geotags of that user in a starting region $\mathcal{R}^s \subset \mathcal{W}$. A \textit{region} $\mathcal{R}$ can refer to either a city or a country from Table~\ref{tab:UserCount}. To evaluate the performance of the location prediction we remove all the geotags of $u_k$ that lie within $\mathcal{R}^t$ and use the geotags of $u_k$ in an other region $\mathcal{R}^s$ to predict the location of the removed data. For this evaluation setup we need users that have visited at least 2 distinct regions. Obviously, when the recommender is operational, recommendations can already be made when a user has visited a single region. 

To build the location similarity model between $\mathcal{R}^s$ and $\mathcal{R}^t$, we first find the most popular locations in these two regions. We use a kernel convolution of the training data with a Gaussian kernel to smoothly cluster the geotags that are near to each other (Section~\ref{sec:1D}). We also find the most important locations per user by computing the kernel convolution over only the user's geotags. Both resulting distributions ($\Phi$, $\Phi_{u_k}$) are combined in the co-occurrence space $\Phi^{CC}$ which estimates the relations between the top locations in both regions (Section~\ref{sec:2D}). The model $\Phi$ will be used to generate a baseline ranking (Section~\ref{sec:Baseline}), the model $\Phi^{CC}$ will be used to predict a personalised location ranking per user (Section~\ref{sec:Res1}-\ref{sec:Res2}).

\section{Peak Finding  ($\Phi$,$\Phi_{u_k}$)}\label{sec:1D}
The geotags of all users are described by the function $f$ which has a Dirac delta pulse at the locations where one of the users created a geotag and zero otherwise:
\begin{equation}
		f(z) = \sum_{l \in \mathcal{L}} \alpha_l \delta(z-l)
\end{equation}
where $\alpha_l$ is a parameter that allows the assignment of different weights per geotag. In this work $\alpha_l$ will be set to 1 for all $l$, other weighting strategies will be discussed in Section~\ref{sec:Discussion}.

We propose to use a Gaussian kernel convolution to obtain a smooth estimate of the density of all photos on the planet $\Phi_{\sigma} = f \ast g_{\sigma}$, where the Gaussian kernel is described by $g_{\sigma}(z) = e^{- \left\|z\right\|^2 / 2\sigma^2}$, for $z \in \mathbb{R}^3$.
%$g(z) = e^{-z^2 / 2\sigma^2}$.
%The kernel convolution is similar to the Parzen density estimation of the data~\cite{Parzen1962}. 
The standard deviation $\sigma$ is used as a scaling parameter (or bandwidth) which gives the opportunity to set the size of the recommended locations. We do not use the common normalisation parameter of a probability density estimation with Gaussian kernels ($1/n\sqrt{2 \pi \sigma^2}$, with $n$ the number of data points) so that the convolution result will directly estimates the total number of photos taken at a certain location instead of the probability. In the rest of this work, we will drop the subscript $\sigma$ for readability. 

In the same way the function describing the geotag profile of a single user $u_k$ is given by: 
\begin{equation}
		f_{u_k}(z) = \sum_{l \in \mathcal{L}_{u_k}} \alpha_l \delta(z-l)
\end{equation}
And the density estimate $\Phi_{u_k} = f_{u_k} \ast g$.

%===============================================================
% BEGIN FIGURE =================================================
\begin{figure}[!t]
\centering
\includegraphics[width=0.7\columnwidth]{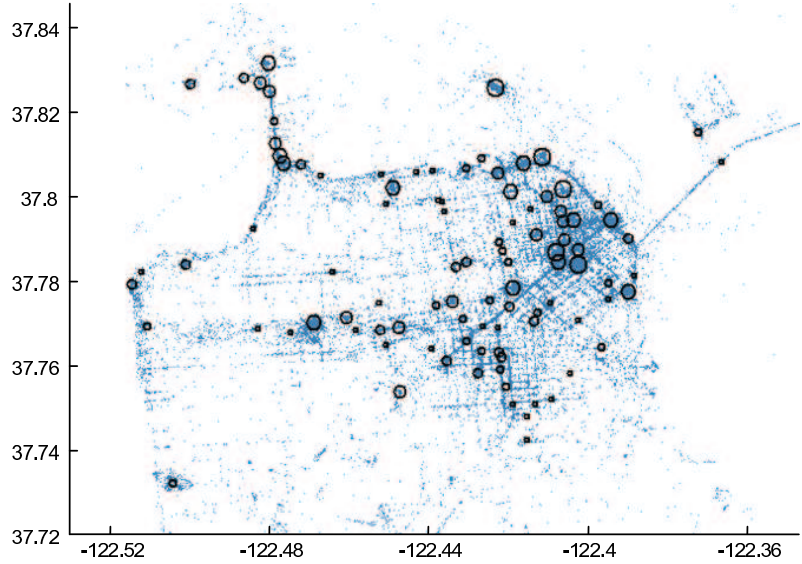}
\caption{The circles indicate the top-100 peaks in San-Francisco at $\sigma=100\,\textup{m}$ where the radius is related to the peak amplitudes. The underlying data points clearly show the structure of the touristic part of the city.}
\label{fig:1DPeaks}
\end{figure}
% END FIGURE ==================================================
%==============================================================

We use $\mathcal{P}$ and $\mathcal{P}_{u_k}$ to denote the local maxima or \textit{peaks} of $\Phi$ and $\Phi_{u_k}$ respectively. These peaks represent the most popular locations for all or a single user. A \textit{mean shift} procedure is used to efficiently find the peaks of the functions~\cite{Cheng1995}. We evaluate the peaks at 19 values of $\sigma$ evenly distributed on a logarithmic scale from 10\,m to 10\,km for cities and 1\,km to 1000\,km on country scale. To ensure that all local maxima are found, we initiate the mean shift procedure with all individual geotags for computation on the finest scale. On each subsequent scale $\sigma$, we use the peaks from the previous scale as seeds. This procedure results in a scale-space that represents the structure of the data and allows us to analyse it at various scales. 

The peaks $p \in \mathcal{P}$, found by the mean shift procedure on all geotags, can now be ranked based on their amplitude to obtain a popularity ranking of the locations in region $\mathcal{R}$ at scale $\sigma$. The application of the mean shift algorithm on geotag data was already proposed by Crandall et al. Compared to their work our scale-space will be more accurate because we use Cartesian coordinates instead of mapping latitude and longitude in a 2D plane~\cite{Crandall2009}. Also, our method differs from Crandall et al.\ as we use a Gaussian kernel instead of a uniform disk. The Gaussian kernel convolution results in a smooth density estimate and does not generate plateau peaks. Other notable similar methods to define a popularity ranking of all locations in a given area are the scale specific clustering in Yahoo!'s World Explorer~\cite{Ahern2007,Rattenbury2007} and the tree-based hierarchical graph used in Microsoft's GeoLife project~\cite{Zheng2009a}. We chose to use the Gaussian scale space as it has a strong theoretical foundation~\cite{Lindeberg1994} and will show to provide a logical solution to the co-occurrence model. 

%===============================================================
% BEGIN FIGURE =================================================
\begin{figure}[!t]
\centering
\includegraphics[width=0.7\columnwidth]{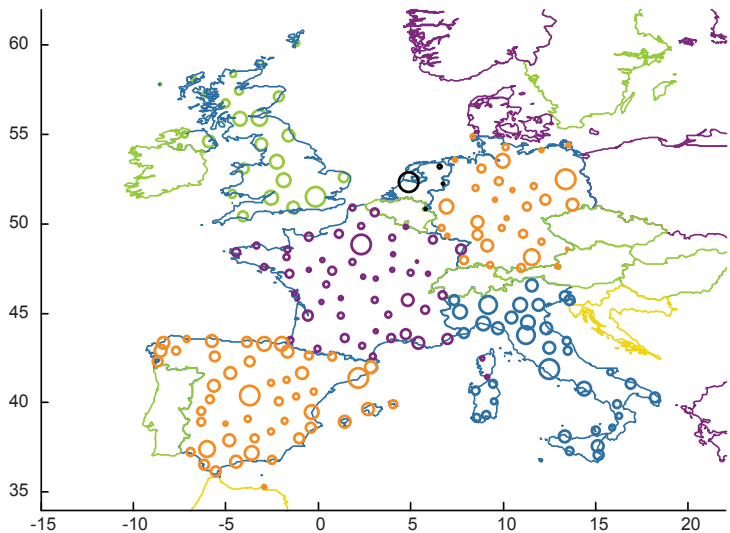}
\caption{The polygons of the European countries in the top-10 most visited (Blue), top-20 (Green), top-30 (Purple) and top-40 (Yellow). The circles indicate the peaks in the top-10 most visited countries with $\sigma=21.5\,\textup{km}$, the radius is related to the peak amplitudes.}
\label{fig:1DPeaksEurope}
\end{figure}
% END FIGURE ==================================================
%==============================================================

In Figure~\ref{fig:1DPeaks} the data points of the training users in the city center of San Francisco are shown (the actual bounding box used in this work is larger). The top-100 peaks with largest amplitude at $\sigma=100\,\textup{m}$ are depicted by circles. The clustering shows that the proposed model does capture most of the well known landmarks like \textit{Alcatraz}, \textit{Union Square Park}, \textit{Coit Tower}, \textit{Yerba Buena Gardens} and \textit{Pier 39}. Long stretched landmarks like the \textit{Golden Gate Bridge}, are not represented by a single cluster but several clusters appear at the popular view points. %We assume that bridges do not represent a specific user interest and therefore allow this flaw...
Figure~\ref{fig:1DPeaksEurope} shows the country polygons in western Europe and for the countries in the top-10 the clusters are shown at a scale of $\sigma=21.5\,\textup{km}$. Most of the main cities are clearly visible on the map. The west of the Netherlands is grouped into a single cluster at this scale, which is reasonable as it is often seen as a single metropolitan area. At smaller scales the individual cities appear.

%===============================================================
% BEGIN FIGURE =================================================
\begin{figure}[!t]
\centering
\includegraphics[width=0.48\columnwidth]{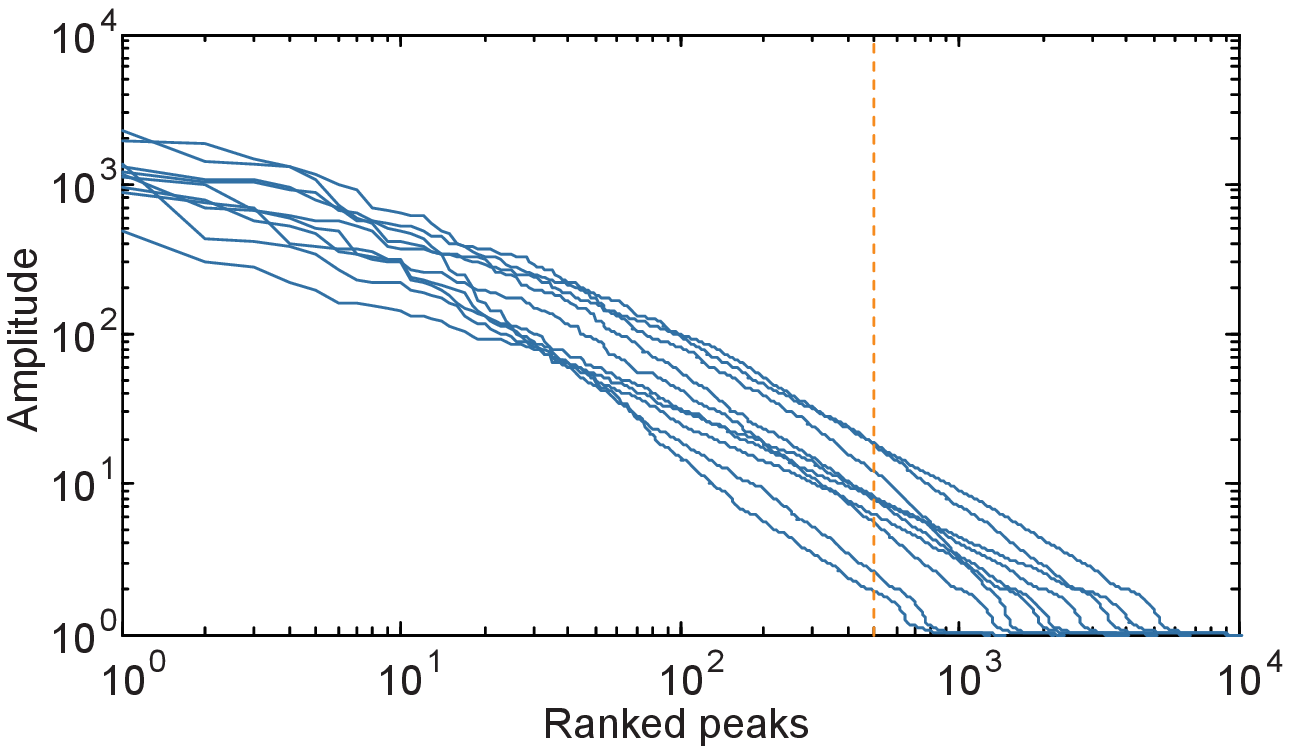}
\includegraphics[width=0.48\columnwidth]{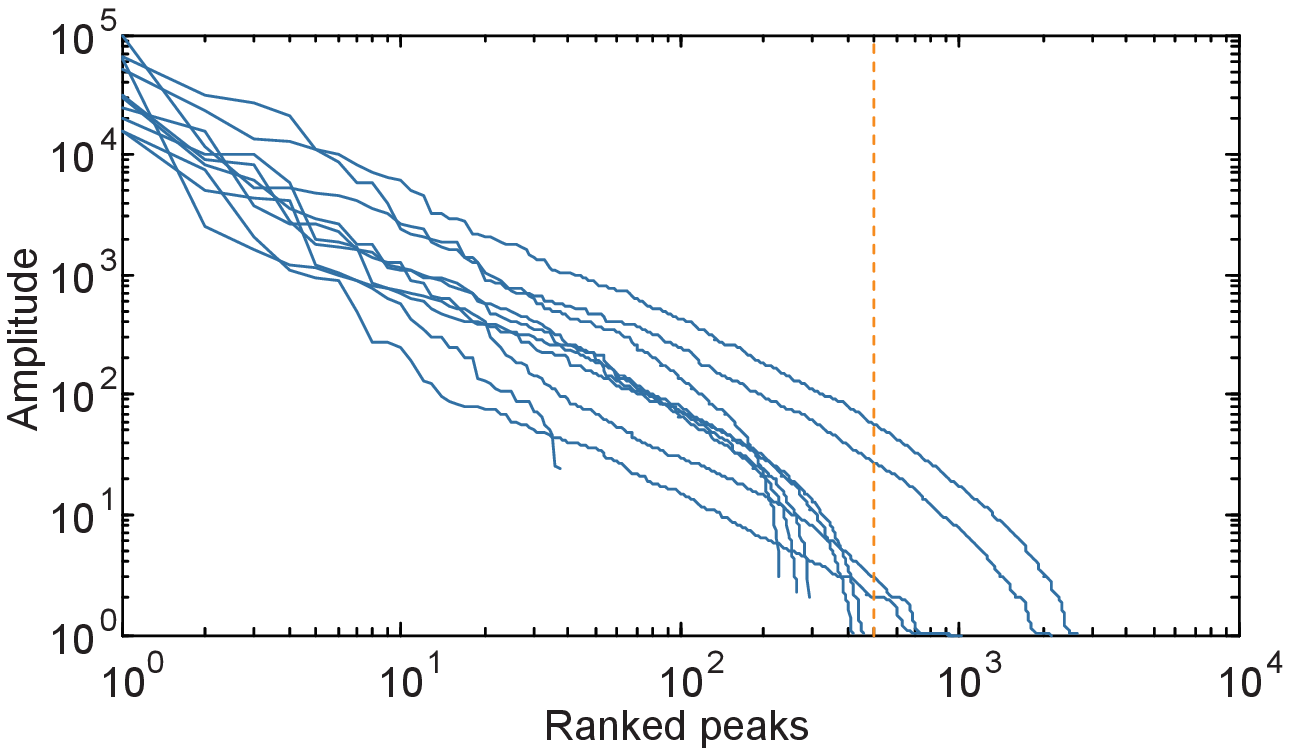}
\caption{The distribution of peak amplitudes at the smallest scales that will be used for evaluation in cities and countries. Left: Each line shows the peak amplitudes in one of the top-10 cities at $\sigma=46\,\textup{m}$. Right: Each line represents one of the top-10 countries at $\sigma=6.8\,\textup{km}$. The dotted lines indicate the cutoff at 500 peaks.}
\label{fig:PeakDist}
\end{figure}
% END FIGURE ==================================================
%==============================================================

For computational efficiency we will only experiment with the top-500 peaks in each region. To check whether we are missing any important peaks in this step we look at the peak amplitude of the 500th peak in Figure~\ref{fig:PeakDist}. As the contribution of each geotag to a peak ranges between 0 and 1, the peak amplitude estimates the number of photos taken there. Because a user can make multiple photos at a single location, the number of users that contribute to the peak will be smaller: users $<$ photos $\approx$ Peak amplitude. 

The values chosen for $\sigma$ will be explained in Section~\ref{sec:Baseline}. At $\sigma = 46\,\textup{m}$ there are only three cities where the 500\supscr{th} peak has an amplitude larger than 10 (London, New York, San Francisco). There are two countries (USA East and USA West) that still have large peaks after the top-500 (Amplitudes: 57 and 28). We believe that a cluster smaller than 10 photos is insignificant for our task and conclude that in most regions no important locations will be lost due to the selection of the top-500 peaks.

\section{Co-occurrence Model ($\Phi^{CC}$)}\label{sec:2D}
%
%
%\subsection{Model computation}\label{sec:2Dmodel}
%
When visiting a country or city, most users actively plan their trip and choose the landmarks to visit based on their interests. Especially, making a photo at a certain location is a clear indication of interest in that location.  Based on these assumptions, we propose to estimate the similarity between two location by the number of users that have made a photo at \textit{both} places. As geotags are continuous points in $\mathcal{W} \subset \mathbb{R}^3$, a method needs to be found that counts the contribution of each of these points to a pair of landmarks. 

\begin{figure}[!t]
%\rput[tl](0.2,1.3) {\large $p^s_m$}
%\rput[tl](0.5,3.6) {\large $\Phi^s$}
%\rput[tl](1.7,0.5) {\large $\Phi^s_{u_2}$}
%\rput[tl](2.3,0.5) {\large $\Phi^s_{u_1}$}
%\rput[tl](6.2,6.6) {\large $p^t_n$}
%\rput[tl](2,5.8) {\large $\Phi^t$}
%\rput[tl](7.8,4.6) {\large $\Phi^t_{u_2}$}
%\rput[tl](7.8,4.0) {\large $\Phi^t_{u_1}$}
%\rput[tl](10.75,2.6) {\large $\Phi^{CC}(c_{m,n})$}
%\rput[tl](10.3,0.4) {\large $c_{m,n}$}
%\rput[tl](9.4,0.5) {\large $c_{u,1}$}
%\rput[tl](10,0.5) {\large $c_{u,2}$}
%\rput[tl](11.1,0.5) {\large $c_{u,3}$}
\begin{minipage}[b]{\columnwidth}%
	\begin{center}
	\includegraphics[width=0.8\columnwidth]{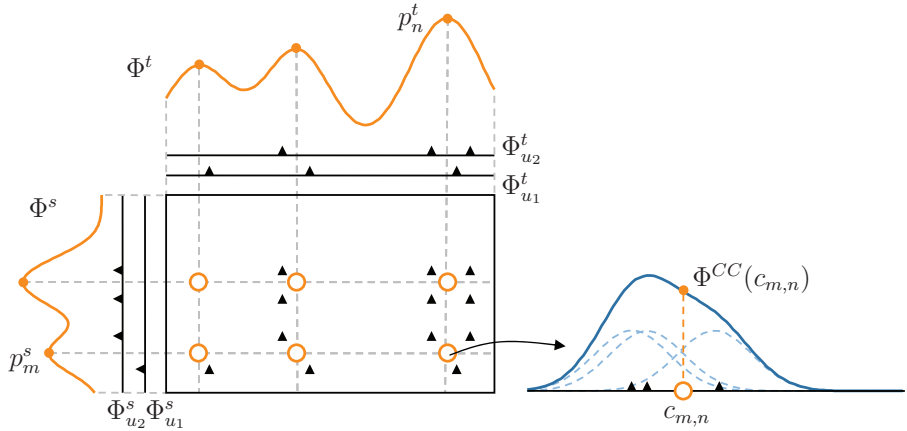}
	\end{center}%
\end{minipage}
\caption{Co-occurrence model. Each user's peaks are mapped into the co-occurrence space (visualised for two users). At the Top-500 peak locations of the prior distribution $\Phi$ the result of the kernel convolution in the co-occurrence space $\Phi^{CC}$ is evaluated. For one point the contribution from the two users is visualised. For visualisation purposes the 6D co-occurrence space is visualised in 2D (left) and 1D (right). %The figure also shows that there might be a small mismatch between the real peak in $\Phi^{CC}$ and the value at the evaluation point $\left\langle l_m,l_n \right\rangle$
}
\label{fig:2DModel}
\end{figure}

We propose to create the location co-occurrence model between two regions $\mathcal{R}^s$ and $\mathcal{R}^t$ as follows. At a chosen scale $\sigma$ the locations visited by $u_k$ are selected by taking his peaks $p_{u_k}^s \in \mathcal{P}_{u_k}^s$ from $\mathcal{R}^s$ and $p_{u_k}^t \in \mathcal{P}_{u_k}^t$ from $\mathcal{R}^t$. The location co-occurrences for this user between the two regions are given by $c_{u_k} \in \mathcal{C}_{u_k}$, where $\mathcal{C}_{u_k} = \left\{ \left\langle  p_{u_k}^s,p_{u_k}^t \right\rangle | p_{u_k}^s \in \mathcal{P}_{u_k}^s, p_{u_k}^t \in \mathcal{P}_{u_k}^t \right\} \subset \mathbb{R}^6$ is the set of all pairwise combinations of this user's peaks in both regions. The points in the co-occurrence space are visualised for two users by the black triangles in Figure~\ref{fig:2DModel}.

When all the peaks of all users are added to this co-occurrence space, the most dense regions represent location pairs that are often visited by the same users, and therefore indicate a strong similarity between the two locations. A smoothed prediction of location similarities can now be derived by computing the kernel convolution over the co-occurrence space, which will be denoted as $\Phi^{CC}$. However, since this space may contain millions of 6 dimensional data points, applying the mean shift algorithm to find the local optima is computationally expensive.

However, the locations of the most prominent landmarks are already known from $\mathcal{P}^s$ and $\mathcal{P}^t$. Therefore we only need to evaluate the value of $\Phi^{CC}$ at the pairwise location combinations from $\mathcal{P}^s$ and $\mathcal{P}^t$, visualised as orange circles in Figure~\ref{fig:2DModel}. For example, when $p_m^s$ and $p_n^t$ are two peaks in $\Phi^s$ and $\Phi^t$ respectively, and the combined location is given by $c_{m,n} = \langle p_m^s,p_n^t \rangle \in \mathbb{R}^6$, the co-occurrence of these two landmarks is defined by the sum over all user contributions:
\begin{equation}
\Phi^{CC}(c_{m,n})=\sum_{u_k \in U} \sum_{c_{u_k} \in \mathcal{C}_{u_k}} e^{-d(c_{m,n},c_{u_k})/2\sigma^2}
\end{equation}
where $d(c_{m,n},c_{u_k})$ is the Euclidean distance between the evaluated landmark combination $c_{m,n}$ and $c_{u_k}$ is a location co-occurrence in the profile of $u_k$. As we have limited the number of peaks per region to 500 there will be maximally 250,000 evaluation points per combination of $\mathcal{R}^s$ and $\mathcal{R}^t$.

The upper left point in the co-occurrence space example in Figure~\ref{fig:2DModel} illustrates that peak intersections from $\Phi^s$ and $\Phi^t$ may exist that do not generate a peak in the co-occurrence space $\Phi^{CC}$: if two locations are simply never visited by a single user, the co-occurrence will be zero. 

We illustrate the computation of $\Phi^{CC}$ at the bottom right evaluation point in Figure~\ref{fig:2DModel}. Three user points contribute significantly to the co-occurrence peak, although also the small contributions from the other peaks are taken into account. The illustration also indicates that the actual peak in the co-occurrence space might be slightly shifted to a different location. The impact of the error introduced by this approximation is discussed in Appendix~\ref{app:full2d}.

%To derive the location similarity model between two regions $\mathcal{R}^s$ and $\mathcal{R}^t$, we first map each user's peaks from $\Phi_{u_k}$ (at a selected $\sigma$) in a co-occurrence space (Visualised in Figure~\ref{fig:2DModel}). For example, when a user has a peak $(x_1,y_1,z_1)$ in $\mathcal{R}^s$ and a peak $(x_2,y_2,z_2)$ in $\mathcal{R}^t$ we create an entry $(x_1,y_1,z_1,x_2,y_2,z_2)$ in the co-occurrence space. A smoothed prediction of location similarities can now be derived by computing the kernel convolution over the co-occurrence space. However, since this space may contain millions of 6 dimensional data points, the mean shift algorithm is computationally expensive. 

%We approximate the result by evaluating the function $\Phi^{CC}$ only at the locations of the top-500 most popular points in both $\mathcal{R}^s$ and $\mathcal{R}^t$ (Max.\;250,000 evaluation points, of which 6 are indicated by the circles in Figure~\ref{fig:2DModel}). For example, when $l_m$ and $l_n$ are two peaks in $\Phi_A$, the relation between them is defined by:
%\begin{equation} 
%\Phi^{CC}(l_m,l_n)=\sum_{\hat{l}_q \in \hat{L}} e^{-d(\left\langle l_m,l_n \right\rangle,\hat{l}_q)/2\sigma^2}
%\end{equation}
%where $\hat{l}_q \in \hat{L}$ is a 6-dimensional point in the co-occurrence space and $d(\left\langle l_m,l_n \right\rangle,\hat{l}_q)$ is the Euclidean distance between the combined point $\left\langle l_m,l_n \right\rangle$ and $\hat{l}_q$. 

%
\section{Results}
\subsection{Baseline Optimisation and Evaluation Criteria}\label{sec:Baseline}
As a baseline, the peaks in $\mathcal{R}^t$ will be ranked on the score determined by the general popularity: $S(p^t_n)=\Phi(p^t_n)$. This results in a static ranking, equal for all users. After ranking the locations, we compute the distance of each of the recommended locations to the nearest peak of the test user in $\mathcal{P}_{u_k}$ (at the same $\sigma$). We then set a threshold $PC$ on this distance and count a recommended location as correct if the nearest of the user's peaks lies within this threshold. At small scale values, many peaks will be predicted close to each other. To make sure the recommender does not get rewarded for the suggestion of a single landmark multiple times, we disqualify a recommended location if it lies within distance $PC$ from an earlier prediction.

The predicted location ranking will be evaluated on four criteria:
\begin{quote}
\textbf{Precision} (P@5), defined as the fraction of correct recommendations in the top-5.

\vspace{1ex}\textbf{Mean average precision} (MAP@50), the mean over the precision values after each correct recommendation in the top-50.

\vspace{1ex}\textbf{NDCG\subscr{IP}}. Similar to Zhou et al.\ we want to express the \textit{surprisal value} of the recommended list in a number~\cite{Zhou2010}. We propose to use the Normalised Discounted Cumulative Gain (NDCG) by J\"arvelin and Kek\"al\"ainen which compares the predicted ranking to the optimal possible ranking~\cite{Jarvelin2002}. The NDCG allows the assignment of a \textit{gain} value to account for differences in relevance between the ranked objects (please refer to~\cite{Jarvelin2002} for details). To measure the surprisal value of the predicted ranking we set the gain of each correctly recommended location $p^t_n$ to the inverse popularity $1/\Phi(p^t_n)$ abbreviated as \textsc{ip}, so that less popular locations contribute more to the result than popular locations. Then we compute NDCG\subscr{IP} over the resulting ranking. The optimal NDCG\subscr{IP} will be obtained when we correctly predict all the user's test locations, but in reverse order of popularity. 

\vspace{1ex}\textbf{Benefit ratio} (BR), the number of users who get an improved recommendation over the baseline divided by the number of users who get a deteriorated recommendation. BR can be computed over any of the previously defined evaluation methods. 
\end{quote}

%===============================================================
% BEGIN FIGURE =================================================
\begin{figure}[!t]
\rput[tl](2.2,3.2) {\footnotesize $32\,\textup{m}$}
\rput[tl](2.5,3.9) {\footnotesize $46\,\textup{m}$}
\rput[tl](2.8,4.55) {\footnotesize $100\,\textup{m}$}
\rput[tl](3.8,5.1) {\footnotesize $147\,\textup{m}$}
\rput[tl](10.0,3.45) {\footnotesize $6.8\,\textup{km}$}
\rput[tl](9.9,4.2) {\footnotesize $21.5\,\textup{km}$}
\rput[tl](11.1,4.55) {\footnotesize $31.6\,\textup{km}$}
\begin{minipage}[b]{\columnwidth}
\begin{center}
\includegraphics[width=0.48\columnwidth]{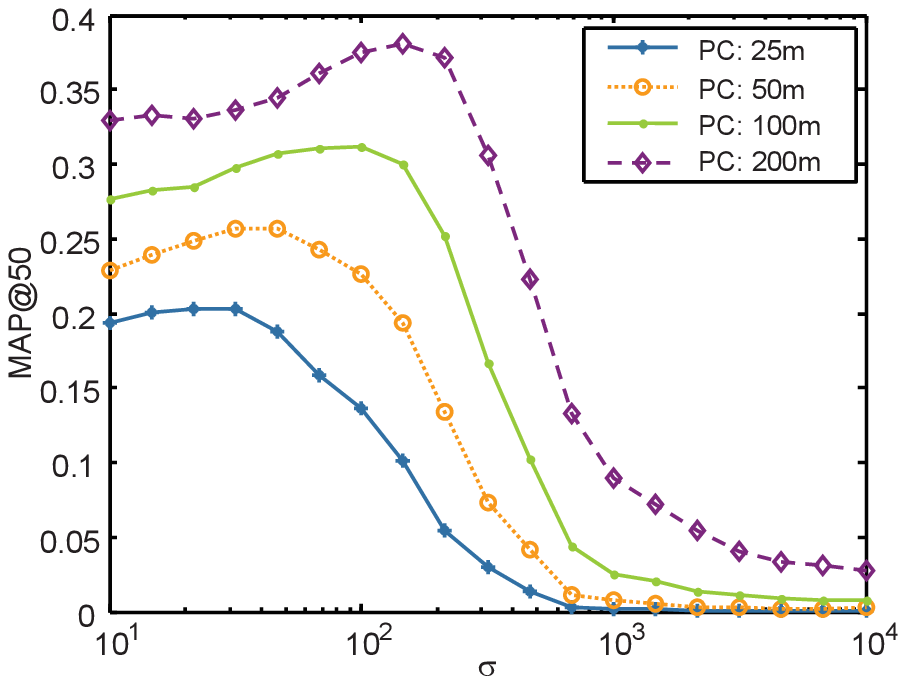}
\hspace{0.01\columnwidth}
\includegraphics[width=0.48\columnwidth]{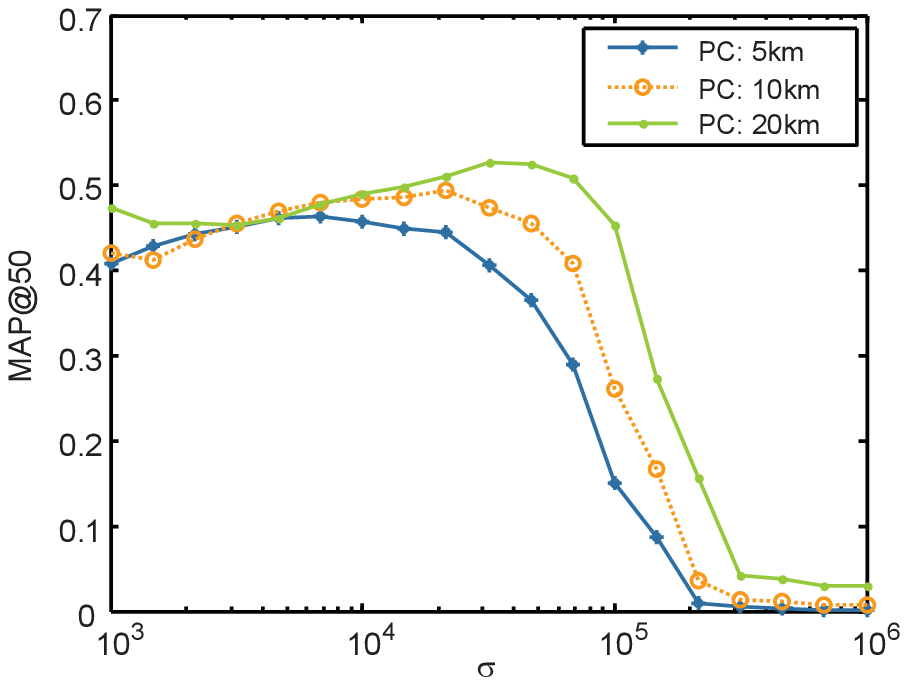}
\end{center}
\end{minipage}
\caption{Performance of the baseline ranking using MAP@50. Left: Results on city scale, for the full range of $\sigma$ and $PC \in \{25,50,100,200\}\,\textup{m}$. Right: Results at country scale for $PC \in \{5,10,20\}\,\textup{km}$.}
\label{fig:Baseline2}
\vspace{-1ex}
\end{figure}
% END FIGURE ==================================================
%==============================================================

To only evaluate users who have provided a decent amount of preference information, we consider those users who have at least 5 peaks at the lowest level of the scale-space ($|\mathcal{P}_{u_k}|\geq5$). At city scale this pruning step means that users must have at least 5 peaks in $\Phi_{u_k}$ at $\sigma=10\,\textup{m}$. At country scale, users need to have at least 5 peaks in $\Phi_{u_k}$ at $\sigma=1\,\textup{km}$.

The optimal $\sigma$ at a chosen value of $PC$ will be estimated based on MAP@50. Compared to P@5, the results on MAP@50 more gradually change with different values of $\sigma$, therefore parameter optimisation on MAP@50 gives a more reliable estimate of the optimal setting. P@5 however gives a more intuitive evaluation on the practical usability of the recommender. We will therefore show the results on both criteria in the next sections. 

In Figure~\ref{fig:Baseline2} the mean MAP@50 is plotted for the baseline ranking for the full range of $\sigma$ values and various settings of $PC$. For all settings, the choice of $\sigma$ has a clear optimum. When $\sigma$ is chosen too small, multiple peaks exist at a single landmark, while for too large $\sigma$ individual landmarks will be missed because they are merged into a single peak. 
At city scale the optimal $\sigma$ is found close to the selected value of $PC$. At country scale we find that the optimal $\sigma$ is larger. This can be explained by the fact that within a city the ratio between the point of interest size and the distance between them is larger than in a country. 
%the objects are often quite close to each other (museums, statues, bridges, etc.), while in a country there is usually a big distance between individual objects (cities, national parks, mountains, etc.). 

At both city and country level, we select two scales for further evaluation. Within city recommendation will be evaluated at $PC=50\,\textup{m}$, $\sigma=46\,\textup{m}$ and $PC=100\,\textup{m}$, $\sigma=100\,\textup{m}$. At country scale we will evaluate recommendations at $PC=5\,\textup{km}$, $\sigma=6.8\,\textup{km}$ and $PC=10\,\textup{km}$, $\sigma=21.5\,\textup{km}$. %http://en.wikipedia.org/wiki/Haplography

\subsection{Recommendation}\label{sec:Res1}
\subsubsection{Generating Recommendations}\label{sec:2Drec}

We compute $\Phi^{CC}(\langle p^s_m,p^t_n \rangle)$ for all paired peaks in the top-500 $p^s_m \in \mathcal{P}^s$ and the top-500 $p^t_n \in \mathcal{P}^t$ in all combinations of $\mathcal{R}^s$ and $\mathcal{R}^t$ (the top-10 cities and countries), based on the set of training users. The derived models can now be used to generate recommendations for the test users. 

As explained in Section~\ref{sec:setup} the geotags of test user $u_k$ in a starting region $\mathcal{R}^s$ will be used to predict the visited locations in $\mathcal{R}^t$. The predicted location ranking in $\mathcal{R}^t$ will then be compared to the locations actually visited by $u_k$. In order to evaluate the performance of the predicted recommendations for a test user, the user therefore needs to have visited at least two distinct regions. In both regions we enforce the pruning settings at $|\mathcal{P}^s_{u_k}| \geq 5  \wedge |\mathcal{P}^t_{u_k}| \geq 5$ as explained in Section~\ref{sec:Baseline}.

\begin{figure}[!t]
%\rput[tl](-0.5,2.5) {\footnotesize $\mathcal{R}^s$}
%\rput[tl](0.1,3.0) {\footnotesize $p^s_{u_k}[3]$}
%\rput[tl](0.1,2.3) {\footnotesize $p^s_{u_k}[2]$}
%\rput[tl](0.1,1.6) {\footnotesize $p^s_m$}
%\rput[tl](0.1,1) {\footnotesize $p^s_{u_k}[1]$}
%\rput[tl](1,4.5) {\footnotesize $\Phi_{u_k}$}
%\rput[tl](4.0,5.0) {\footnotesize $\mathcal{R}^t$}
%\rput[tl](3.6,4.6) {\footnotesize $p^t_n$}
%\rput[tl](4.4,2.7) {\footnotesize $d(p^s_m,p^s_{u_k}[3])$}
%\rput[tl](1.6,1.9) {\footnotesize $d(p^s_m,p^s_{u_k}[2])$}
%\rput[tl](4.4,1.2) {\footnotesize $d(p^s_m,p^s_{u_k}[1])$}
%\rput[tl](9.5,2.8) {\footnotesize $\Phi^{CC} (c_{m,n})$}
%\rput[tl](9.4,1.4) {\footnotesize $\sigma$}
%\rput[tl](10.7,2.0) {\footnotesize $\Phi^{CC} (c_{m,n}) e^{-d(p^s_m,p^s_{u_k}[2])/2 \sigma^2}$}
%\rput[tl](8.8,0.5) {\footnotesize $p^s_{u_k}[1]$}
%\rput[tl](10.2,0.5) {\footnotesize $p^s_{u_k}[2]$}
%\rput[tl](10.2,1.1) {\footnotesize $p^s_m$}
%\rput[tl](11.1,0.5) {\footnotesize $p^s_{u_k}[3]$}
%\rput[tl](10.4,0.6) {\footnotesize $d(p^s_m,p^s_{u_k}[1])$}
%\rput[tl](10.4,0.6) {\footnotesize $\left\|l_m-l_{k_1}\right\|$}
\begin{minipage}[b]{\columnwidth}
\centering
\noindent\makebox[\textwidth]{%Will center the image
\includegraphics[width=0.8\columnwidth]{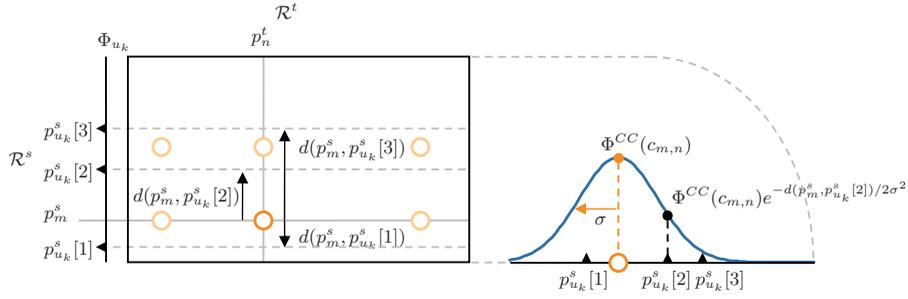}}
\end{minipage}
\caption{Computing recommendations with the location co-occurrence model. For each peak $p^s_m$ in $\mathcal{R}^s$ all contributions of all the user's geotags are aggregated using a Gaussian distribution as weight function. Then the final score of a location $p^t_n$ in $\mathcal{R}^t$ is derived from the sum over all $p^s_m$.}
\label{fig:Recommend}
\end{figure}

The score of location $p^t_n$ in $\mathcal{R}^t$ for user $u_k$ is now derived by:
\begin{equation}\label{eq:rec}
S^{CC}(p^t_n,u_k)=\sum_{p^s_m \in \mathcal{P}^s} \sum_{p^s_{u_k} \in \mathcal{P}^s_{u_k}} \Phi^{CC} (\langle p^s_m,p^t_n \rangle) e^{-d(p^s_m,p^s_{u_k})/2\sigma^2}
\end{equation}
which counts the contribution of each of the user's peaks $p^s_{u_k}$ in $\mathcal{R}^s$ to each of the landmarks $p_m^s$ in $\mathcal{R}^s$, and weights each of these landmarks with the co-occurrence model. To predict the recommendations for $u_k$ when traveling to $\mathcal{R}^t$, the locations $p^t_n$ are ranked according to this score and the top ranked locations are recommended. This computation is visualised for a user $u_k$ with three geotags in $\mathcal{R}^s$ in Figure~\ref{fig:Recommend}.

%===============================================================
% BEGIN FIGURE =================================================
% END FIGURE ==================================================
%==============================================================

%
\subsubsection{Recommendation performance}
\begin{table}[!b]
\caption{Results of the baseline ($S$) compared to the recommender ($S^{CC}$), for two scales at both city and country level.}
\label{tab:Results1}
\centering
\footnotesize
\begin{tabular}{lcccccccc}
\toprule
&\multicolumn{4}{c}{City}&\multicolumn{4}{c}{Country}\\
&\multicolumn{2}{c}{$PC=50\,\textup{m}$}&\multicolumn{2}{c}{$PC=100\,\textup{m}$}&\multicolumn{2}{c}{$PC=5\,\textup{km}$}&\multicolumn{2}{c}{$PC=10\,\textup{km}$}\\
&\multicolumn{2}{c}{$\sigma=46\,\textup{m}$}&\multicolumn{2}{c}{$\sigma=100\,\textup{m}$}&\multicolumn{2}{c}{$\sigma=6.8\,\textup{km}$}&\multicolumn{2}{c}{$\sigma=21.5\,\textup{km}$}\\
&$S$&$S^{CC}$&$S$&$S^{CC}$&$S$&$S^{CC}$&$S$&$S^{CC}$\\
\cmidrule(r{.25em}){2-3}\cmidrule(lr{.25em}){4-5}\cmidrule(lr{.25em}){6-7}\cmidrule(l){8-9}
P@5             &	0.237 & 0.237 &	0.293 & 0.300 &	0.266 & 0.274 &	0.257 & 0.261\\
MAP@50          &	0.311 & 0.312 &	0.370 & 0.377 &	0.437 & 0.445 &	0.482 & 0.488\\
NDCG\subscr{IP} & 0.237 & 0.238 &	0.272 & 0.277 &	0.287 & 0.293 &	0.358 & 0.365\\
BR-P@5      &\multicolumn{2}{c}{1.034}&\multicolumn{2}{c}{1.375}&\multicolumn{2}{c}{1.419}&\multicolumn{2}{c}{1.337}\\
BR-MAP@50   &\multicolumn{2}{c}{1.046}&\multicolumn{2}{c}{1.246}&\multicolumn{2}{c}{1.248}&\multicolumn{2}{c}{1.298}\\
BR-NDCG\subscr{IP}&\multicolumn{2}{c}{1.108}&\multicolumn{2}{c}{1.361}&\multicolumn{2}{c}{1.292}&\multicolumn{2}{c}{1.476}\\
\bottomrule
\end{tabular}
\end{table}

We now compare the ranking on $S$ to the ranking predicted by $S^{CC}$. Table~\ref{tab:Results1} contains the results at the two selected scales for between-city and between-country recommendation. The presented values are averaged over all possible recommendations for all city/country pairs in the top-10 lists. At city scale the results are based on 16,620 measurements, with an average user size of 9 locations (median). At country scale we can evaluate 13,476 recommendations, with a median user size of 7.

For all settings and all evaluation methods our model improves over the baseline. We test the significance of the improvement using a Wilcoxon signed rank test, which tests the hypothesis that the difference between the matched samples in the two sets comes from a distribution with zero median. At a confidence level of 1\% only the results on P@5 for $\sigma = 46\,\textup{m}$ are not significant. Probably too many landmarks will be represented by multiple peaks at this scale, making the co-occurrence model less accurate. 

The improved results on NDCG\subscr{IP} indicate that not only the rank position of the test results improves, but also the surprisal value of the presented recommendations. The co-occurrence model gives better performance while less popular locations are observed at the top of the ranking. This shows that the method correctly learns how the preference of the user differs from the average. 

Although BR shows a decent improvement when the recommendation model is used, the mean absolute improvement on the individual evaluation criteria is small. For many users the popularity based baseline and the personalised ranking of recommended locations are very similar. Two reasons can be given for these small differences. First, many users do not have a single preference (e.g.\;only visit botanical gardens), but visit many types of landmarks when they come to a new location. With the proposed co-occurrence model, the combined recommendations based on these mixed preference profiles converge to the prior ranking. Second, because many people visit the most popular locations in the target region the evaluation method actually expects us to recommend these. This is inherent to the evaluation of recommendations with a train and test set. 

In Section~\ref{sec:Res2} we will see that when a single type of landmark is used as starting location and we manually asses the recommended locations, the prediction is highly accurate and we can use more extreme weighting methods to exploit the location co-occurrence. %The difference between these two evaluation methods will be more thoroughly discussed in Section~\ref{sec:Discussion}.

\subsubsection{Tourist Filter}

We hypothesize that people who visit both $\mathcal{R}^s$ and $\mathcal{R}^t$ for touristic purposes will benefit more from the recommendations than people who actually live in one of the cities. To confirm this hypothesis we implement a tourist filter as follows: Based on the creation date of the photos in the \Flickr\ data a user qualifies as tourist in a certain city if all his photos in that city are taken in $n$ periods of 14 days. So in the 3x14 filter we allow the user to visit a single city 3 times, and all the user's photos have to be taken in at most 3 different windows of 14 days. 

The results with three different tourist filters applied in both $\mathcal{R}^s$ and $\mathcal{R}^t$ are presented for $\sigma=100\,\textup{m}$ in Table~\ref{tab:ResultsTourist}. First, we observe that both the baseline and the recommendation performance go up when a more stringent filter is used. So tourists conform more to the overall visiting behaviour than city inhabitants. Second, when we set a more strict tourist filter, the performance difference between the recommender and the baseline goes up. This means that touristic behaviour in one city should be predicted by touristic behaviour in another city. 

Table~\ref{tab:ResultsTourist} also indicates the number of recommendations (\textit{Recs}) that can be evaluated with each filter. We need a user two have made a touristic visit in at least two different cities in order to evaluate the performance. These two criteria cause the number of evaluations to drop quite quickly.  

\begin{table}[!t]
\caption{Results of city scale recommendation at $\sigma=100\,\textup{m}$ for different tourist filters. The best results are obtained for the most strict filter (1x14).}
\label{tab:ResultsTourist}
\centering
\footnotesize
\begin{tabular}{lcccccccc}
\toprule
&\multicolumn{8}{c}{City, $PC=100\,\textup{m}$, $\sigma=100\,\textup{m}$}\\
\cmidrule{2-9}
&\multicolumn{2}{c}{All}&\multicolumn{2}{c}{3x14}&\multicolumn{2}{c}{2x14}&\multicolumn{2}{c}{1x14}\\
&$S$&$S^{CC}$&$S$&$S^{CC}$&$S$&$S^{CC}$&$S$&$S^{CC}$\\
\cmidrule(r{.25em}){2-3}\cmidrule(lr{.25em}){4-5}\cmidrule(lr{.25em}){6-7}\cmidrule(l){8-9}
P@5             &	0.293 & 0.300 &	0.321 & 0.330 &	0.331 & 0.339 &	0.339 & 0.351\\
MAP@50          &	0.370 & 0.377 &	0.409 & 0.417 &	0.419 & 0.427 &	0.430 & 0.440\\
NDCG\subscr{IP} & 0.272 & 0.277 &	0.301 & 0.308 &	0.307 & 0.314 &	0.318 & 0.325\\
BR-P@5            &\multicolumn{2}{c}{1.375}&\multicolumn{2}{c}{1.511}&\multicolumn{2}{c}{1.491}&\multicolumn{2}{c}{1.687}\\
BR-MAP@50         &\multicolumn{2}{c}{1.246}&\multicolumn{2}{c}{1.338}&\multicolumn{2}{c}{1.358}&\multicolumn{2}{c}{1.422}\\
BR-NDCG\subscr{IP}&\multicolumn{2}{c}{1.361}&\multicolumn{2}{c}{1.491}&\multicolumn{2}{c}{1.547}&\multicolumn{2}{c}{1.614}\\
Recs &\multicolumn{2}{c}{16,620}&\multicolumn{2}{c}{8576}&\multicolumn{2}{c}{6600}&\multicolumn{2}{c}{3536}\\
\bottomrule
\end{tabular}
\end{table}

\begin{table}[!b]
	\caption{Results on recommendation of the locations for the last day of a city visit.}
  \label{tab:InCity}
	\centering
  \scriptsize
	\begin{tabular}{lcccc}
	\toprule
	&\multicolumn{4}{c}{City, $\sigma = 100\,\textup{m}$, $PC=100\,\textup{m}$ }\\
	\cmidrule{2-5}
	&\multicolumn{2}{c}{No pruning} &\multicolumn{2}{c}{$|\mathcal{P}^{t}_{u_k}|\geq5$}\\
	&$S$&$S^{CC}$&$S$&$S^{CC}$\\
	\cmidrule{2-3}\cmidrule{4-5}
	P@5             &	0.042 & 0.047 &	0.108 & 0.129\\
	MAP@50          &	0.099 & 0.119 &	0.197 & 0.244\\
	NDCG\subscr{IP} & 0.126 & 0.141 &	0.208 & 0.234\\
	BR-P@5      &\multicolumn{2}{c}{2.452} &\multicolumn{2}{c}{5.182}\\
	BR-MAP@50   &\multicolumn{2}{c}{1.966} &\multicolumn{2}{c}{2.690}\\
	BR-NDCG\subscr{IP}&\multicolumn{2}{c}{1.982} &\multicolumn{2}{c}{2.531}\\
	Recs &\multicolumn{2}{c}{18,344} &\multicolumn{2}{c}{896}\\
	\bottomrule
	\end{tabular}\\
\end{table}

\subsubsection{Within-City Recommendation}
Song et al.\ showed that the every day mobility patterns of people are highly predictable 93\% of the time~\cite{Song2010}. Other related work on location prediction also focused on making recommendations close to the current location of a user~\cite{Horozov2006,Takeuchi2006,Zheng2010}. We suspect that prediction of touristic behaviour in previously unvisited areas is a much harder task. First, touristic behaviour is less predictable than every day life behaviour. Second, remote predictions allow many more possibilities than nearby recommendations. 

To test whether we can use our model for within-city recommendations we compute the co-occurrence space $\Phi^{CC}$ within each city ($\mathcal{R}^t = \mathcal{R}^s$) and set the self co-occurrence of each location to 0. For each user $u_k$ in the test set that has been to $\mathcal{R}^t$, we cut off the last day of photos made in that city. We use the geotags created by $u_k$ on all previous days as starting points and try to predict the user's behaviour on the final day of his stay. To split the user's data in days we use the creation date and time of the photos shifted backwards by 4.5 hours based on the results in Figure~\ref{fig:DataTime}. 

Table~\ref{tab:InCity} gives the results at $\sigma = 100\,\textup{m}$ averaged over all users (\textit{No pruning}), and limited to users who have at least 5 peaks in $\mathcal{P}^t_{u_k}$ at this scale in both the test day and the training days. The absolute evaluation scores are lower than the scores reported in between-city recommendation, because we have fewer evaluation points in this setup. After pruning, the median user has 6 points on the test day, compared to a median of 9 in city to city recommendation. 

The relative improvement with the personalised model is much larger for within-city recommendation than that for between-city recommendation. Especially for users with many geotags on the training and test day the personalised prediction clearly outperforms the baseline. Unfortunately, only few users (\textit{Recs}) have provided enough data to pass the pruning settings. These findings indicate that adapting the location prediction to a user's personal interest is easier if the user stays within the same city. 

\begin{figure}[!t]
	\centering	\includegraphics[width=0.45\columnwidth]{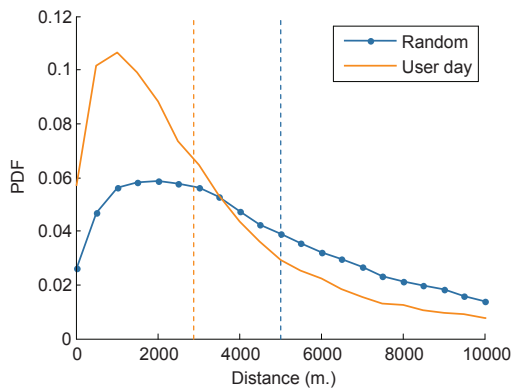}
  \caption{PDF of distance between two \textit{random} geotags and between the last and previous days of a single user (\textit{user day}). }
  \label{fig:DayDist}
\end{figure}

We assume that the reason for this improvement is that users have a bias to make many photos within a certain area (e.g.\;close to the hotel). To verify this, we plot the probability density function (PDF) of the distance between two randomly selected geotags and the PDF of the distance between a geotag selected from the training days and a geotag selected from the test day of a single user (Figure~\ref{fig:DayDist}). The dotted lines indicate the median of both distributions. Clearly the geotags selected from a single user have a larger probability to be close together. This location prior explains why recommendations within a single region are easier to predict than between two remote locations, confirming the second intuition given above, that remote locations allow more possibilities than nearby ones.

\subsubsection{Conclusions}
%For many users the co-occurrence model can give improved recommendations, the absolute improvement is however small because users have mixed preferences. 
Because many users visit the same popular locations, prediction according to the prior travel probability is hard to improve upon. Although the absolute improvement is small, the co-occurrence model can give improved recommendations for most users.

Tourists can be selected by setting a maximum value on the number of days spent on a certain location. We find that tourists comply more with the general travel preference and are therefore more easy to predict by the baseline. Also, the relative improvement of the personalised model over the baseline is larger than for the average user, which shows that tourists have a clear preference that relates their behaviour in different cities. This shows that the location co-occurrence model based on the travel history of tourists can effectively be used to predict personalised travel recommendations. We have used a simple tourist filter and suggest that more elaborate methods could be used based on the users' profile information or textual tags. 

Within-city recommendations are easier because the training data contains a location prior. If we know where the user was in the past few days, he is more likely to be in the same place the next day. 

\subsection{Serendipitous Ranking}\label{sec:Res2}
\subsubsection{Ranking Criteria}
Using part of the users' real data points as test set, we have evaluated whether we can predict where the user will go if he is not influenced by a recommender. This evaluation is however strongly biased by the most popular locations in the target area. As most people will visit the \textit{Eiffel Tower} when they get to Paris, it pays off to predict this with the recommender. However, the user would benefit more from a recommendation of a location that is not obvious and perhaps even unknown to the user. Related work on recommender systems has therefore argued that manual judgement of the recommended items is necessary for the evaluation of novel recommendations~\cite{Celma2008}. 

To test whether the proposed co-occurrence model can be used to produce serendipitous recommendations, we have manually annotated various sets of landmarks at city and country scale. We first use one of the landmarks ($p_m^s$) in $\mathcal{R}^s$ as starting point and try to predict the annotated landmarks ($p_n^t$) that fall in the same category in $\mathcal{R}^t$, using the following known ranking criteria: 
\begin{quote}
\textbf{Prior ($S$)} Ranking based on $\Phi(p_n^t)$.

\vspace{1ex}\textbf{Direct ($S^{CC}$)} As the user profile now consists of only a single peak from $\Phi$ in $\mathcal{R}^s$, Equation~\ref{eq:rec} reduces to a ranking based directly on $\Phi^{CC}(c_{m,n})$.

\vspace{1ex}\textbf{Cosine (CS)} Ranking based on $\Phi^{CC}(c_{m,n}) / \sqrt{\Phi(p_m^s)\Phi(p_n^t)}$. Cosine similarity corrects for the popularity bias by dividing the co-occurrence by the popularity of both individual landmarks. 
\end{quote}

We also propose a new ranking method, which assigns the prior amplitudes ($\Phi$) as weight to all locations and then compares the weight difference between the initial and new ranking:
\begin{quote}
\textbf{RankDiff (RD)} Let $R_1$ be the rank index (position in the ranked list) of a location based on $\Phi(p_n^t)$ and $R_2$ the rank index of the same location in $\Phi^{CC}(c_{m,n})$. Let $\Psi$ be the list of peak amplitudes of $\Phi$ ranked in descending order. RankDiff is now defined as $RD(p_n^t) = \Psi(R_2)-\Psi(R_1)$.
\end{quote}
The rationale behind this method is that a location that used to be at rank $R_1$ and had an amplitude of $\Phi(p_n^t)$ managed to reach a new ranking of $R_2$ where a location with amplitude $\Phi(p_x^t)$ used to be. The difference between these two amplitudes can now be seen as the amount of evidence needed to accomplish this rank gain. 

Note that we also considered other ranking algorithms, that performed worse or very similar to any of the above (i.e.\;\textit{Jaccard coefficient}, \textit{Pointwise Mutual Information (PMI)}, \textit{Lift}~\cite{Tan2002,Ohsaki2005}); the results of these ranking criteria are therefore left out of the discussion.

\begin{table}[!t]
\caption{Baseball stadium set. The prior rank is the rank index based on $S$ in the corresponding region.}
\label{tab:BaseballSet}
\centering
\footnotesize
\begin{tabular}{lcccc}
\toprule
Stadium & City & Prior Rank & Longitude & Latitude\\
\cmidrule{1-5}
Yankee Stadium & New York &	27 & 40.8271 & -73.9281\\
City Field &	New York & 44 & 40.7557 & -73.8481\\
Richmond Co. Bank Ballpark & New York & 151 & 40.6457 & -74.0761\\
AT\&T Park &	San Francisco &	13 & 37.7785 & -122.3896\\
Dodger Stadium &	Los Angeles &	12 & 34.0735 & -118.2400\\
Nationals Park &	Washington & 22 & 38.8729 & -77.0076\\
Wrigley Field &	Chicago & 5 & 41.9479 & -87.6558\\
Cellular Field &	Chicago &	18 & 41.8300 & -87.6340\\
\bottomrule
\end{tabular}
\end{table}

\begin{table}[!t]
\caption{Modern art museum set. The prior rank is the rank index based on $S$ in the corresponding region.}
\label{tab:MuseumSet}
\centering
\footnotesize
\begin{tabular}{lcccc}
\toprule
Museum & City & Prior Rank & Longitude & Latitude\\
\cmidrule{1-5}
Tate Modern & London & 4 & 51.5081 & -0.0990\\
Museum of Modern Art & New York & 5 & 40.7610 & -73.9771\\
Guggenheim Museum & New York & 12 & 40.7831 & -73.9591\\
Centre Pompidou & Paris & 6 & 48.8604 & 2.3520\\
Hirshhorn Museum & Washington & 10 & 38.8888 & -77.0230\\
MACBA & Barcelona & 7 & 41.3832 & 2.1668\\
Fundacio Miro & Barcelona & 28 & 41.3686 & 2.1597\\
Neue Nationalgalerie & Berlin & 23 & 52.5070 & 13.3681\\
Haus der Kulturen der Welt & Berlin & 21 & 52.5187 & 13.3648\\
Hamburger Bahnhof Museum & Berlin & 28 & 52.5283 & 13.3719\\
\bottomrule
\end{tabular}
\end{table}

\begin{table}[!b]
\caption{Results on baseball stadium and modern art prediction. Mark that the number of test locations in all cities is small, therefore the maximum possible P@5 equals 0.30 for baseball stadiums and 0.32 for modern art museums.}
\label{tab:BaseballRes}
\centering
\footnotesize
\begin{tabular}{lcccccccccc}
\toprule
&\multicolumn{5}{c}{Baseball}&\multicolumn{5}{c}{Modern Art}\\
\cmidrule(r{.25em}){2-6}\cmidrule(l{.25em}){7-11}
Method & Up & Down & P@5 & R@5 & P@R & Up & Down & P@5 & R@5 & P@R\\
\cmidrule(r{.25em}){1-1}\cmidrule(r{.25em}){2-6}\cmidrule(l{.25em}){7-11}
$S$ & 0 & 0 & 0.04 & 0.09 & 0 & 0 & 0 & 0.07 & 0.26 & 0\\
$S^{CC}$ & 45 & 3 & 0.16 & 0.58 & 0.29 & 53 & 18 & 0.10 & 0.41 & 0.19\\
CS & 41 & 7 & 0.15 & 0.47 & 0.24 & 30 & 49 & 0.07 & 0.27 & 0.06\\
RD & 44 & 4 & 0.23 & 0.76 & 0.46 & 39 & 38 & 0.12 & 0.43 & 0.25\\
\bottomrule
\end{tabular}
\end{table}
\subsubsection{City scale}
We manually annotate a set of baseball stadiums (Table~\ref{tab:BaseballSet}) and a set of modern or contemporary art venues (Table~\ref{tab:MuseumSet}) in the top-10 cities. 
We now repeatedly select one of the cities as $\mathcal{R}^t$ and rank all landmarks in that region based on one landmark in $\mathcal{R}^s$. As evaluation we compute the number of times a target location (from one of the two sets) goes \textit{up} or \textit{down} in the ranking compared to a ranking based on $S$, the precision at 5 (P@5), recall at 5 (R@5), defined as the fraction of correct results ranked in the top-5 and precision at R (P@R), where R is the total number of correct locations that can be recommended. For all evaluations the peaks in $\Phi$ at $\sigma=100\,\textup{m}$ are used, since at this scale it is easy to manually relate each peak to a single landmark.

The results in Table~\ref{tab:BaseballRes} show that almost all baseball stadiums are related to each other as 45 out of 48 times a stadium gets a higher ranking based on co-occurrence than on the prior (48 is the total number of possible ways to select two landmarks from different cities). A ranking directly based on $S^{CC}$ does get the target locations higher in the list, but the more popular locations are often still at the very top of the ranking, resulting in a limited P@5, R@5 and P@R. The other methods make more mistakes on up/down, but RankDiff clearly improves precision and recall. The P@R of 0.46 indicates that RankDiff gets the target stadium(s) to the very top of the ranking in about half of the cases, which is a remarkable achievement given the relatively low prior rank of the stadiums.

Further inspection of the ranking produced by cosine similarity shows that many very small peaks are ranked at the top. Cosine similarity can easily generate a high score when an unfamiliar starting location is used, if by coincidence the users who have been there have also another location in common. RankDiff is somewhat more conservative as it is more dependent on the absolute value of $\Phi^{CC}$ than the relative difference between $\Phi^{CC}$ and $\Phi$.

\begin{figure}[!t]
\rput[tl](2.7,9.2) {\large Baseball}
\rput[tl](10.0,9.2) {\large Modern art}
\begin{minipage}[b]{0.41\columnwidth}%
\centering
%\vspace{1ex}
\includegraphics[height=1.74in]{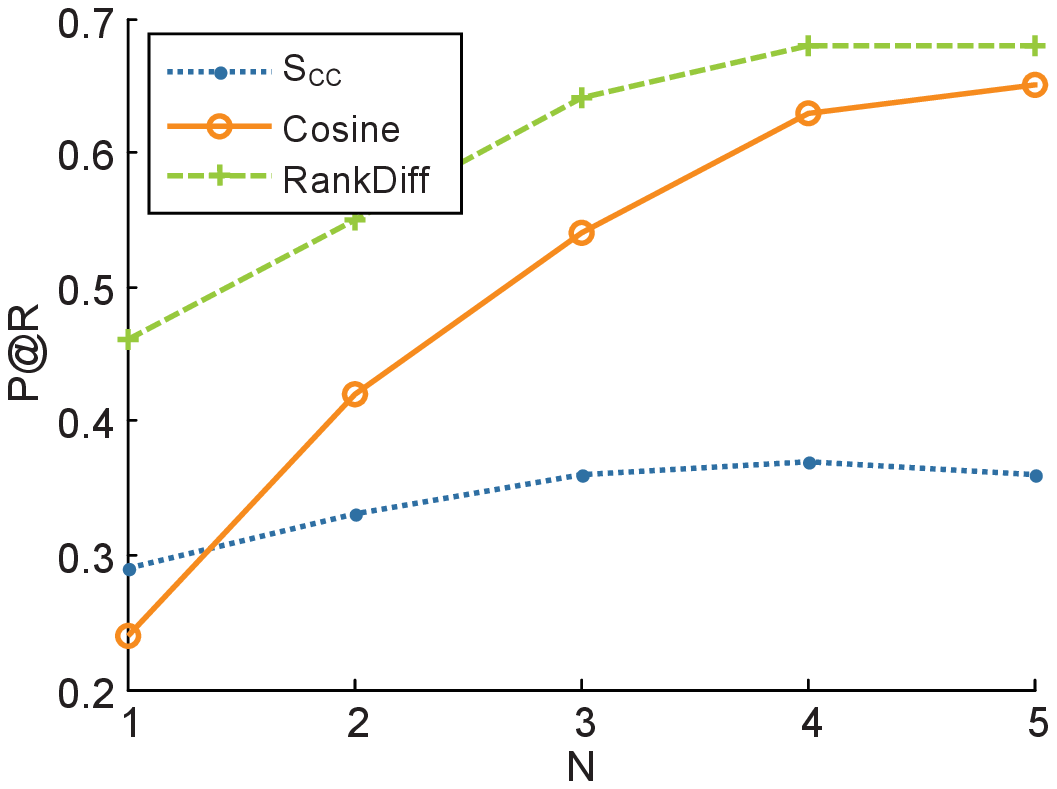}
\hspace{0ex}
\includegraphics[height=1.75in]{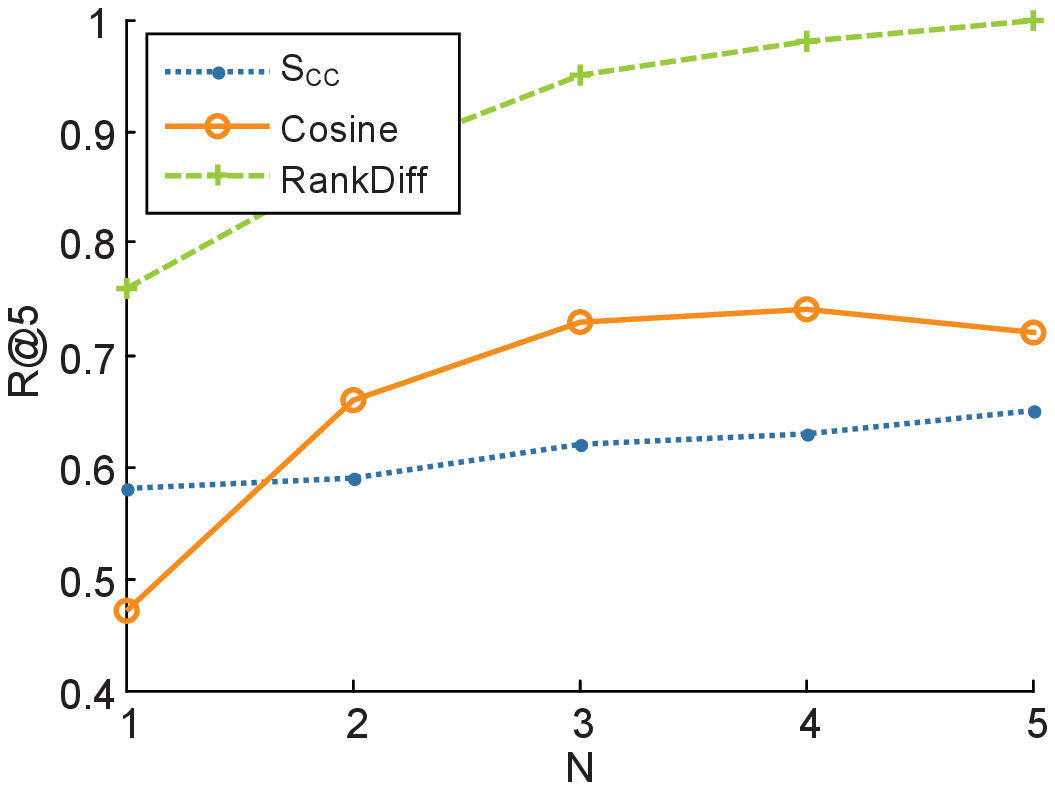}
\end{minipage}
\hspace{2pt}
\begin{minipage}[b]{0.55\columnwidth}%
\centering
%\vspace{1ex}
\includegraphics[height=1.74in]{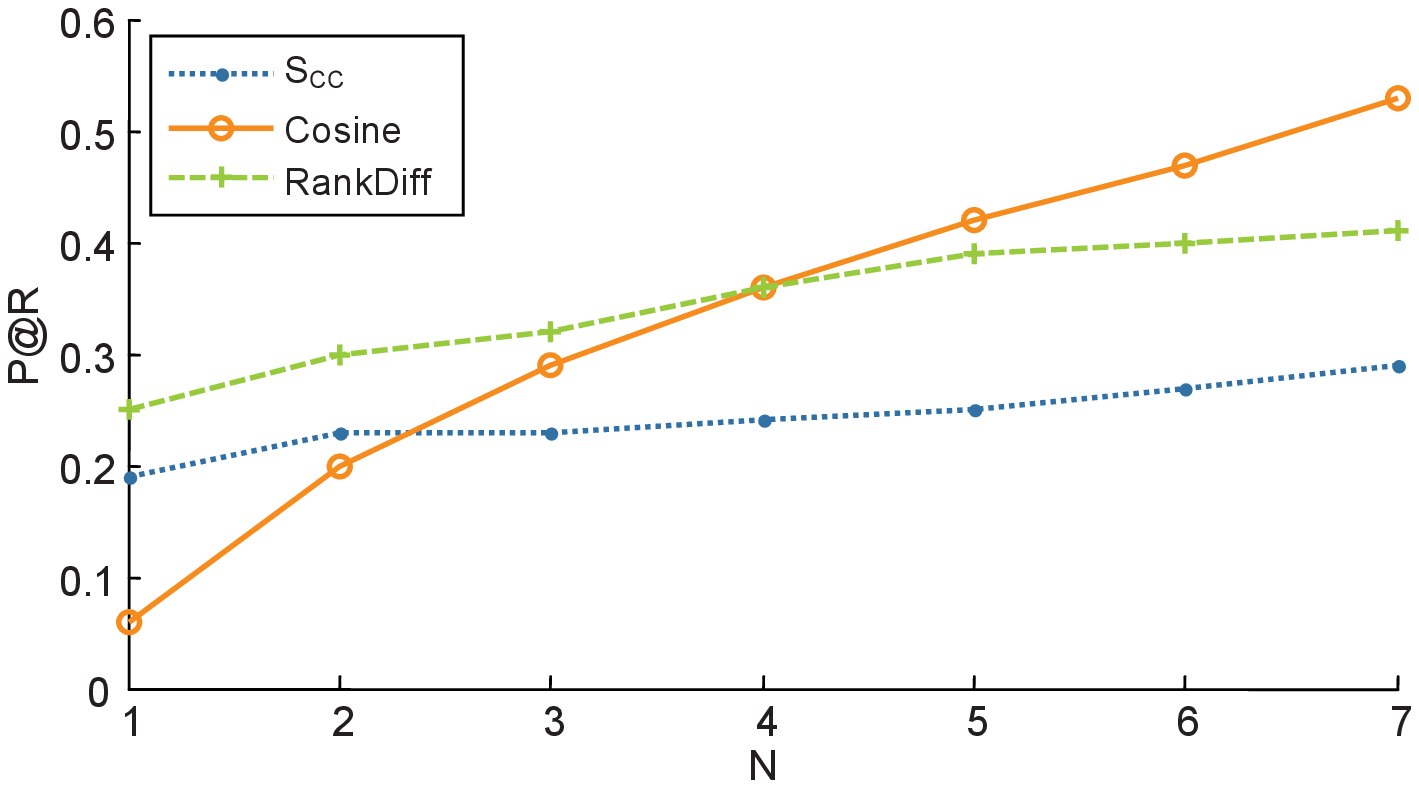}
\hspace{0ex}
\includegraphics[height=1.75in]{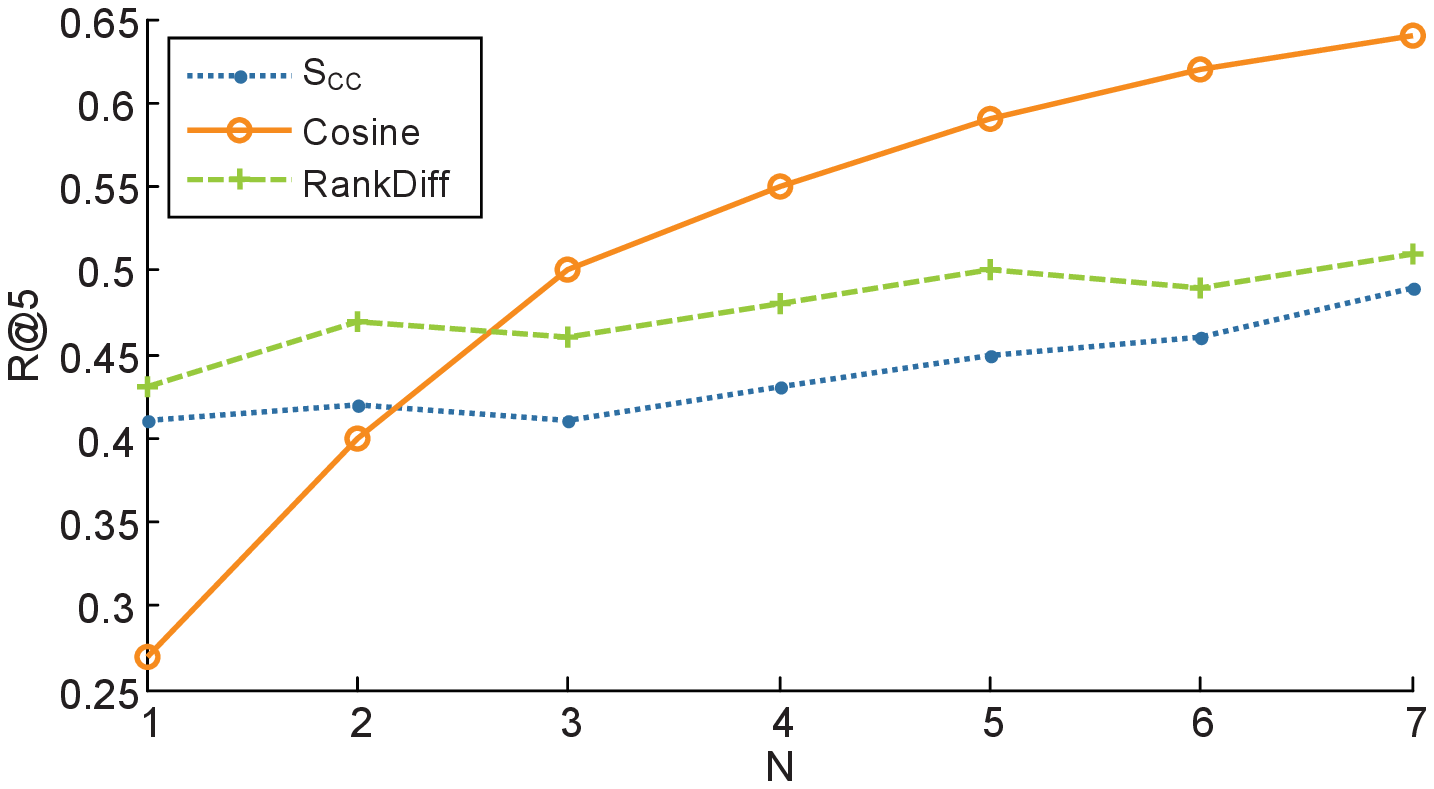}
\end{minipage}
\caption{Performance difference of $S^{CC}$, Cosine and RankDiff with increasing profile information ($N$). The two left panels show the P@R and R@5 for Baseball recommendation, the right panels for Modern art.}
\label{fig:Ninc}
\vspace{-0.5ex}
\end{figure}

Although the modern art data set appears to be less coherent, the order of the methods is similar. Because many of the venues already have a high prior ranking it is hard to improve the prediction. RankDiff again gives the best performance on precision and recall.

To study the benefit of having more profile information from a user, Figure~\ref{fig:Ninc} shows P@R and R@5 for the three personalised methods on both data sets while the number of starting locations is increased. When two starting locations are located in different cities we simply sum the $\Phi^{CC}$ values before computing the ranking criteria. The results are averaged over all target cities and all possible combinations of $N$ landmarks selected from the other cities. 

When the recommendations from more starting points are aggregated the prediction generally gets better. The prediction of baseball stadiums based on RankDiff even reaches a R@5 of 1, meaning that in all cases the target locations are ranked in the top-5. If more information is present, Cosine similarity is less prone to mistakes and shows a steep upward trend in performance.

\subsubsection{Country scale}
To evaluate the co-occurrence model at country scale, we manually annotate a large set of the peaks in USA West at $\sigma = 21.5\,\textup{km}$ and select various starting locations in other countries to see how they influence the ranking in USA West. 

Based on the prior ranking (not shown) the top-10 of locations in USA West contain 9 cities and only 1 national park (\textit{Yosemite NP}). If we use \textit{Ayers Rock} in Australia as starting point we expect recommendations that refer more to natural locations and less to cities. A ranking directly based on $S^{CC}$ does show that some natural parks increase their ranking, but the co-occurrence with the top-4 cities is still larger, simply because their prior visit probability is larger (see Table~\ref{tab:Ayers}). 

\begin{figure}[!t]
\rput[c](2.1,4.1) {\small Query: Ayers Rock}
\rput[c](6.7,4.1) {\small Painted Hills (1)}
\rput[c](11.1,4.1) {\small Craters of the Moon (2)}
\rput[c](2.1,0.4) {\small Monument Valley (3)}
\rput[c](6.7,0.4) {\small Bryce Canyon (5)}
\rput[c](11.1,0.4) {\small Mount Rushmore (6)}
\centering
\includegraphics[width=0.89\columnwidth]{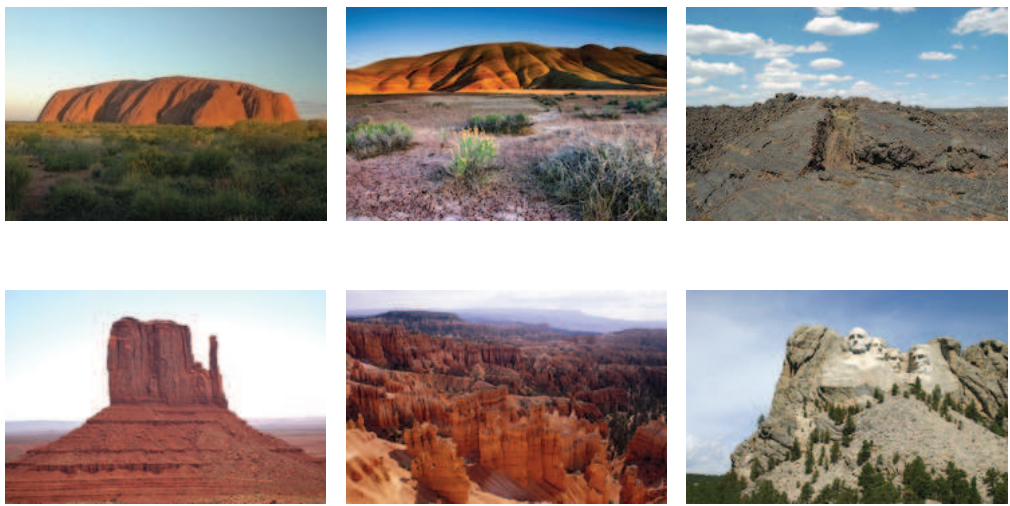}
\caption{When Ayers Rock in Australia is used as query, the top recommendations in USA West contain many famous rock formations.}
\label{fig:Ayers}
\end{figure}

We find that especially cosine similarity returns very interesting recommendations. Figure~\ref{fig:Ayers} and Table~\ref{tab:Ayers} show that almost all places in the top-10 refer to rock formations in the USA, which is quite amazing since absolutely no semantic information (like textual tags) is used in the prediction. 

In this example, cosine similarity seems to give better results than RankDiff. On this scale there are hardly any obscure peaks, therefore we can take the risk of using a method that can get small peaks very high in the ranking, and cosine similarity is able to get peaks from the lower part of the ranking to the top. This introduces more risk in the recommender, but can also give more interesting and serendipitous recommendations.

\begin{table}[!t]
\caption{Top-10 recommendations based on Ayers Rock, Australia. R is the new ranking, PR is the prior ranking (based on $\Phi$).}
\label{tab:Ayers}
\centering
\footnotesize
\noindent\makebox[\textwidth]{%Will center the image
\begin{tabular}{lllllll}
\toprule
&\multicolumn{2}{c}{$S^{CC}$}&\multicolumn{2}{c}{Cosine}&\multicolumn{2}{c}{Rankdiff}\\
\cmidrule(r{.25em}){2-3}\cmidrule(rl{.25em}){4-5}\cmidrule(l{.25em}){6-7}
R & PR & Location & PR & Location & PR & Location\\
\cmidrule{1-7}
1 & 1 & San Fransisco & 129 & Painted Hills SP & 4 & Las Vegas \\
2 & 4 & Las Vegas & 122 & Craters of the Moon NM & 32 & Bryce Canyon NP\\
3 & 3 & Los Angeles & 44 & Monument Valley SP & 44 & Monument Valley SP\\
4 & 2 & Seattle & 99 & Idaho Falls & 36 & Mt. Rushmore NM\\
5 & 32 & Bryce Canyon NP & 32 & Bryce Canyon NP & 13 & Lake Tahoe \\
6 & 44 & Monument Valley SP & 36 & Mt. Rushmore NM & 14 & Grand Canyon NP\\
7 & 5 & Portland & 62 & Mt. Shasta & 17 & Maui\\
8 & 36 & Mt. Rushmore & 49 & Crater lake & 49 & Crater lake NP \\
9 & 13 & Lake Tahoe & 141 & Roswell & 62 & Mt. Shasta\\
10 & 14 & Grand Canyon NP & 153 & Socorro / Box Canyon & 122 & Craters of the Moon NM\\
\bottomrule
\end{tabular}}
\end{table}

\subsubsection{Conclusions}
When the co-occurrence model is used to generate a location ranking based on a single preference point, we observe great performance increase over the prior ranking. A ranking based on $S^{CC}$ directly does get the correct locations higher in the list, but not to the very top of the ranking. We find that more extreme weighting methods can be used to fully exploit the co-occurrence model. 

Cosine similarity can give very small peaks as recommendations when the co-occurrence happens to be relatively large compared to the prior visiting probability. The Ayers rock example showed that this can give very interesting results. Using solely the location history of \Flickr\ users, we were able to relate rock formations on completely opposite sides of the world. 

When limited information is available the risk of recommending something unknown is high when cosine similarity is used. The proposed method \textit{RankDiff} is more conservative, the results are more reliable but may be less surprising. On a manually annotated set of baseball stadiums we showed that the RankDiff method is able to perfectly predict where a stadium in an unvisited city is located if several other stadiums are used as starting points.

%===============================================================
% BEGIN FIGURE =================================================
% END FIGURE ==================================================
%==============================================================

%
%\section{Discussion}\label{sec:Discussion}
%

%
\section{Conclusions and Discussion}\label{sec:Conclusions}\label{sec:Discussion}
We have proposed to approximate the Gaussian kernel convolution over the co-occurrence space of \Flickr\ geotags to obtain a location similarity model. This new approach to predict recommendations in a continuous object space can effectively be used to recommend locations matching a user's preference. Recommendations can be made close to the location of the user, so that we can suggest landmarks for the next day on a city visit. More interesting, the co-occurrence model can be used to make recommendations in a previously unvisited city or country which is useful while planning a holiday. The bandwidth of the Gaussian kernel controls the size of the target locations, which allows application at a scale of choice (city and country level in this work).
The results suggest that recommendations based on the co-occurrence model are both more accurate and more surprising than a ranking based on the prior travel probability. A simple filter to distinguish inhabitants from tourists indicates that touristic behaviour is more informative for the prediction of a user's behaviour in another city. %Making recommendations in the same city as where the user has been in the last few days is more easy compared to distant recommendations because places in the direct vicinity of the user can be recommended. 

In this work we have set the weight of all geotags equal, but the proposed model can deal with differently valued data points. We discussed the choice to ignore the number of photos in batch uploads, but a weighting method could be proposed to integrate this information in the amplitude of the data point. Furthermore, the importance of a photo could be estimated on external information sources like the textual tags or the interestingness ranking used by \Flickr.

%Because different landmarks have different sizes fixing $\sigma$ for the evaluation can never be optimal for all landmarks. Computing the full scale-space over the location co-occurrences would allow to find the most significant peaks at each scale and combine them in a single recommendation. 

By filtering the set of geotags on the \textit{accuracy} value in the \Flickr\ database we have selected only geotags that are accurate on street level, thereby losing about 40\% of the original data. One could argue whether this accuracy filter is necessary if predictions are made on a larger scale (e.g.\;between-country recommendation). 
The function that describes a set of geotags is in this work defined as a collection of Dirac delta pulses. To integrate the geotag accuracy into this function, it naturally follows that each geotag could itself be described by a Gaussian distribution, where the standard deviation is dependent on the accuracy. In this way inaccurate geotags do not influence predictions on small scale, but do contribute on larger scales. 

Recommendation evaluation with a training and test set has a drawback. Because of the strongly skewed prior travel distribution most of the locations in a user's test set are well-known popular places. These places will dominate the parameter optimisation of the model, resulting in a personalised model that does not differ much from the prior ranking. The popular locations are however not the most interesting places to recommend, because the user is probably already familiar with them or can easily find them in regular travel guides. 

To really evaluate whether a recommender gives interesting, user specific recommendations, manual assessments are inevitable. Using manually annotated locations on both city and country scale we have shown that more strict ranking methods can be used to produce more serendipitous recommendations. A ranking based on cosine similarity can give very interesting and novel recommendations, but also has the possibility of recommending something irrelevant based on data noise. The proposed RankDiff method is more conservative but gives stable good recommendations in all experiments. Based on these results we can assume that these weighting methods will also be more effective in a recommendation system, when the full user profile is used as training data.

% Appendix
\appendices
\section{APPENDIX: Full 6D kernel convolution}\label{app:full2d}
\setcounter{section}{1}

As indicated in the model description in Section~\ref{sec:2D}, the computation of the co-occurrence model at the prior peak locations is an approximation of the real peaks in the co-occurrence model. To estimate the error introduced by this approximation we have used the mean-shift algorithm to compute the peaks of the full Gaussian kernel convolution on the 6D co-occurrence space for the city pair Berlin-Barcelona at $\sigma = 100\,\textup{m}$.

We compare the top-50 similarity relations generated by both methods in the co-occurrence space between Berlin and Barcelona. Using manual evaluation, we find that 44 out of 50 relations uniquely refer to the same landmarks. The median distance of the top-50 peaks in our approximation to the nearest peak in the full convolution is $26\,\textup{m}$. The measured peak amplitude at the landmark locations will always be smaller than the nearest peak in the full convolution. We find that the average decay in peak amplitude in the approximation is -2.4\%. 

The small differences between both models show that the approximation proposed in this work can effectively be used to predict the most co-occurring locations between two cities.  

%If we match all peaks to this nearest peak in the other method, the median difference in rank index equals 2 and the average decay in peak amplitude in the approximation is -2.4\%.

% Acknowledgments
%\begin{acks}
%The authors would like to thank ... (Jeroen de ridder?)
%\end{acks}


\end{document}